\documentstyle[11pt,aaspp4]{article}

\newcommand{\bq}{\begin{equation}}
\newcommand{\eq}{\end{equation}}

\newcommand{\kcorr}{$K$-correction}
\newcommand{\kcorrs}{$K$-corrections}

\begin{document}
\def\refitem{\par\parskip 0pt\noindent\hangindent 20pt}
\normalsize
 
\title{The Farthest Known Supernova: Support for an Accelerating Universe
and a Glimpse of the Epoch of Deceleration\footnote{Based on observations with
the NASA/ESA {\it Hubble Space Telescope}, obtained at the Space Telescope
Science Institute, which is operated by AURA, Inc., under NASA contract NAS
5-26555}\\ {\it Accepted to the Astrophysical Journal}}
\vspace*{0.3cm}

 Adam G. Riess\footnote{Space Telescope Science Institute, 3700 San Martin
Drive, Baltimore, MD 21218}, Peter E. Nugent\footnote{Lawrence Berkeley
National Laboratory, Berkeley, CA 94720}, Brian P. Schmidt\footnote{Mount
Stromlo and Siding Spring Observatories, Private Bag, Weston Creek P.O. 2611,
Australia}, John Tonry\footnote{Institute for Astronomy, University of Hawaii,
2680 Woodlawn Dr., Honolulu, HI 96822}, Mark Dickinson$^2$, Ronald L. Gilliland$^2$, Rodger
I. Thompson\footnote{Steward Observatory, University of Arizona, Tucson, AZ
85721}, Tam\'as Budav\'ari\footnote{Department of
Physics and Astronomy, The Johns Hopkins University, Baltimore, MD 21218; and
Department of Physics, E\"{o}tv\"{o}s University, Budapest, Pf.\ 32, Hungary,
H-1518}, Stefano Casertano$^{2}$, Aaron S. Evans\footnote{Department of Physics
and Astronomy, SUNY, Stony Brook, NY 11794-3800}, Alexei V. Filippenko\footnote{Department of Astronomy, University of
California, Berkeley, CA 94720-3411}, 
Mario Livio$^2$, David B. Sanders$^5$, Alice E. Shapley\footnote{Palomar
Observatory, California Institute of Technology, Mail Code 105-24, Pasadena, CA
91125}, Hyron Spinrad$^7$, Charles C. Steidel$^{10}$, Daniel Stern\footnote{Jet
Propulsion Laboratory, California Institute of Technology, Mail Code 169-327,
Pasadena, CA 91109}, Jason Surace\footnote{SIRTF Science Center, California
Institute of Technology, Mail Code 314-6, Pasadena, CA 91125}, and Sylvain
Veilleux\footnote{Department of Astronomy, University of Maryland, College
Park, MD 20742}

\begin{abstract}

We present photometric observations of an apparent Type Ia supernova (SN~Ia) at
a redshift of $\sim$1.7, the farthest SN observed to date.  The supernova, SN
1997ff, was discovered in a repeat observation by the {\it Hubble Space
Telescope} ({\it HST}) of the Hubble Deep Field--North (HDF-N), and
serendipitously monitored with NICMOS on {\it HST} throughout the Thompson et
al. GTO campaign.  The SN type can be determined from the host galaxy type: an
evolved, red elliptical lacking enough recent star formation to provide a
significant population of core-collapse supernovae.  The classification is
further supported by diagnostics available 
from the observed colors and temporal
behavior of the SN, both of which match a typical SN Ia.  The
photometric record of the SN includes a dozen flux measurements in the $I$,
$J$, and $H$ bands spanning 35 days in the observed frame.  The redshift
derived from the SN photometry, $z=1.7\pm0.1$, is in excellent agreement with
the redshift estimate of $z=1.65\pm0.15$ derived from the
$U_{300}B_{450}V_{606}I_{814}J_{110}J_{125}H_{160}H_{165}K_s$ photometry of the
galaxy.  Optical and near-infrared spectra of the host provide a very tentative
spectroscopic redshift of 1.755.  Fits to observations of the SN provide
constraints for the redshift-distance relation of SNe~Ia and a powerful test of
the current accelerating Universe hypothesis.  The apparent SN brightness is
consistent with that expected in the decelerating phase of the preferred
cosmological model, $\Omega_M \approx 1/3, \Omega_\Lambda \approx 2/3$. It is
inconsistent with grey dust or simple luminosity evolution, candidate
astrophysical effects which could mimic previous evidence for an accelerating
Universe from SNe Ia at $z \approx 0.5$.  We consider several sources of
potential systematic error including gravitational lensing, supernova
misclassification, sample selection bias, and luminosity calibration errors.
Currently, none of these effects alone appears likely to challenge our
conclusions.  Additional SNe~Ia at $z > 1$
will be required to test more exotic alternatives to the accelerating
Universe hypothesis and to probe the nature of dark energy.

\end{abstract}
subject headings:  supernovae: general --- cosmology: observations

\section{Introduction} 

The unexpected faintness of Type Ia supernovae (SNe Ia) at $z \approx 0.5$
provides the most direct evidence that the expansion of the Universe is
accelerating, propelled by ``dark energy'' (Riess et al.  1998; Perlmutter et
al. 1999).  This conclusion is supported by measurements of the characteristic
angular scale of fluctuations in the cosmic microwave background (CMB) which
reveal a total energy density well in excess of the fraction attributed to
gravitating mass (de Bernardis et al. 2000; Balbi et al. 2000; Jaffe et
al. 2001).

However, contaminating astrophysical effects can imitate the evidence for an
accelerating Universe.  A pervasive screen of grey dust could dim SNe Ia with
little telltale reddening apparent from their observed colors (Aguirre 1999a,b;
Rana 1979, 1980).  Although the first exploration of a distant SN Ia at
near-infrared (near-IR) wavelengths provided no evidence of nearly grey dust,
more data are needed to perform a definitive test (Riess et al. 2000).

A more familiar challenge to the measurement of the global acceleration or
deceleration rate is luminosity evolution (Sandage \& Hardy 1973).  The lack of
a complete theoretical understanding of SNe Ia and an inability to identify
their specific progenitor systems undermines our ability to predict with
confidence the direction or degree of luminosity evolution (H\"{o}flich,
Wheeler, \& Thielemann 1998; Umeda et al. 1999a,b; Livio 2000; Drell, Loredo,
\& Wasserman 2000; Pinto \& Eastman 2000; Yungelson \& Livio 2000).  The weight
of empirical evidence appears to disfavor evolution as an alternative to dark
energy, as the cause of the apparent faintness of SNe Ia at $z \approx 0.5$
(see Riess 2000 for a review).  However, the case against evolution remains
short of compelling.

   The extraordinary claim of the existence of dark energy requires a high level
of evidence for its acceptance.  Fortunately, a direct and definitive test is
available.  It should be possible to discriminate between cosmological models
and ``impostors'' by tracing the redshift-distance relation to redshifts
greater than one.

\subsection{The Next Redshift Octave and the Epoch of Deceleration}

If the cosmological acceleration inferred from SNe Ia is real, it commenced
rather recently, at $0.5 < z < 1$.  Beyond these redshifts, the Universe was
more compact and the attraction of matter dominated the repulsion of dark
energy.  At $z > 1$ the expansion of the Universe should have been decelerating
(see Filippenko \& Riess 2000).  The observable result at $z \geq 1$ would be
an apparent {\it increased} brightness of SNe Ia relative to what is expected
for a non-decelerating Universe.  However, if the apparent faintness of SNe Ia
at $z \approx 0.5$ is caused by dust or simple evolution, SNe Ia at $z > 1$
should appear fainter than expected from decelerating cosmological models.
More complex parameterizations of evolution or extinction which can match both
the accelerating and decelerating epochs of expansion would require a
higher order of fine tuning and are therefore less plausible.

Measuring global deceleration at $z > 1$ provides additional cosmological
benefits.  To constrain the equation of state of dark energy (and distinguish a
cosmological constant from the decaying scalar fields described by the
``quintessence'' hypothesis; Peebles \& Ratra 1988; Caldwell, Dav\'e, \&
Steinhardt 1998), it is necessary to break degeneracies which exist between the
global densities of mass and dark energy.  Observations of SNe Ia in this next
redshift octave are well suited to deciphering the nature of dark energy,
motivating recent proposals to develop a wide-field optical and near-IR space
mission (Curtis et al. 2000; Nugent 2000).  To better determine the
merits and technical requirements of such a mission, it will be important to
closely study the first few SNe detected at these redshifts. In addition, the
study of SNe~Ia at $z > 1$ can provide meaningful constraints on progenitor
models (Livio 2000; Nomoto et al. 2000; Ruiz-Lapuente \& Canal 1998), after
surveys of such SNe over a range of high redshifts are completed.

   Both the Supernova Cosmology Project (SCP; Perlmutter et al. 1995) and the
High-z Supernova Search Team (HZT; Schmidt et al. 1998) have pursued the
discovery of SNe Ia in this next redshift interval.  In the fall of 1998 the
SCP reported the discovery of SN 1998ef at $z=1.2$ (Aldering et al. 1998).  The
following year the HZT discovered a SN Ia at $z=1.2$ (SN 1999fv) as well as at
least one more at $z \approx 1.05$ (Tonry et al. 1999; Coil et al. 2000).
These data sets, while currently lacking the statistical power to discriminate
between cosmological and astrophysical effects, are growing, and in the future
may provide the means to break degeneracies.

   In early 1998, Gilliland \& Phillips (1998) reported the detection of two 
SNe,
SN 1997ff and SN 1997fg, in a reobservation of the Hubble Deep Field--North
(HDF-N) with WFPC2 through the F814W filter.  The elliptical host of SN 1997ff
indicated that this supernova was ``most probably a SN Ia...[at] the greatest
distance reported previously for SNe,'' but the observations at a single epoch
and in a single band were insufficient to provide useful constraints on the SN
and hence to perform cosmological tests (Gilliland, Nugent, \& Phillips 1999,
hereafter GNP99).

   Here we report additional, {\it serendipitous} observations of SN 1997ff
obtained in the Guaranteed Time Observer (GTO) NICMOS campaign (Thompson et
al. 1999) and in General Observer (GO) program 7817 (Dickinson et al. 2001), as
well as spectroscopy of the host.  The combined data set provides the ability
to put strong constraints on the redshift and distance of this supernova and
shows it to be the highest-redshift SN~Ia observed (to date).  These
measurements further provide an opportunity to perform a new and powerful test
of the accelerating Universe by probing its preceding epoch of deceleration.

   In \S 2 of this paper, we describe the observations of the SN and its host
in the HDF-N and report photometry of the SN from the NICMOS campaign.  In \S 3
we analyze the observations to constrain the SN parameters: redshift, 
luminosity,
age of discovery, and distance.  The constraints are used to extend the
distance-redshift relation of SNe Ia to $z > 1$ and to discriminate between
cosmological models and contaminating astrophysical effects. Section 4 contains
a discussion of the systematic uncertainties in our measurements and their
implications.  We summarize our findings in \S 5.

\section{Observations}
\subsection{The Discovery of SN 1997ff}

   Between Dec. 23 and Dec. 26, 1997, Gilliland \& Phillips (GO 6473)
re-observed the HDF-N with {\it HST} to detect high-redshift SNe.  These
observations were obtained with WFPC2 (F814W) during 18 {\it HST} orbits in the
continuous viewing zone (CVZ) and at a spacecraft orientation as closely
matched to the original HDF-N as possible.  To critically sample the WF
point-spread function (PSF) and robustly reject all hot pixels, CCD defects,
and noise fluctuations, the exposures were well dithered using 18 different
sub-pixel and multi-pixel offsets.  Additional F300W frames were obtained
during the bright portion of the CVZ orbits to support improved rejection of
transient hot pixels.  The total F814W exposure time in the second HDF-N epoch
(6300~s) was 51\% of that obtained in the original epoch 2.0 years prior.

    After careful processing, the second epoch was registered with the first
and difference frames in both temporal directions were produced.  Robust SN
detection thresholds were determined by a Monte Carlo exercise of adding PSFs
of varying brightness onto host galaxies of varying redshifts.  From this
exercise it was determined that a brightness threshold of $m_I < 27.7$
(Johnson-Cousins) would ensure the rejection of all spurious transients.
Simulations of completeness indicated that 95\% of SNe at $m_I=27$ coincident
with host galaxies at $1.5 < z < 1.9$ would be discovered.  Only transients
which were brighter than the rejection threshold, visible in each of three
subsets and near a host galaxy, were identified as SNe.  Candidates coincident
with hosts' centers were discarded as possible active galactic nuclei.  The
harvest from the HDF-N SN search was two robust SN detections: SN 1997fg and SN
1997ff at a signal-to-noise ratio ($S/N$) of 20 and 9, respectively.  The
former was hosted by a late-type galaxy with a spectroscopic redshift of
0.95.

  SN 1997ff was discovered at $m_I=27.0$ mag, RA = 12$^h$36$^m$44.11$^s$, Dec
=+62$^\circ$12$^\prime$44.8$^{\prime\prime}$ (equinox J2000),
0.16$^{\prime\prime}$ southwest of the center of the host galaxy, 4-403.0
(Williams et al. 1996).  The host has been classified as an elliptical galaxy
based on measurements of its surface-brightness profile, concentration,
asymmetry, and colors as well as by visual inspection (Williams et al. 1996;
Fernandez-Soto, Lanzetta, \& Yahil 1999; Dickinson 1999; Thompson et
al. 1999; Budav\'ari et al. 2000).  A section of the HDF-N near the SN host is
shown in Figure 1 as observed with WFPC2 and NICMOS.  GNP99 favored the
classification of SN 1997ff as Type Ia due to the red, elliptical host.

Photometric redshift determinations of the host had been published using only
$U_{300}B_{450}V_{606}I_{814}$ photometry of the HDF-N ($z=0.95$; Sawicki, Lin,
\& Yee 1997), as well as from the later addition of $J_{125}H_{165}K_s$ data
from the ground ($z=1.32$; Fernandez-Soto et al. 1999). To span
the rest-frame optical breaks in the spectral energy distribution (SED) of
galaxies with $z > 1$ and reliably estimate their photometric redshift, it is
necessary to employ both optical and near-IR data.  GNP99 assumed the
Fernandez-Soto et al. (1999) redshift which employed the best available
coverage of the host SED to date.  However, even with the monochromatic
detection of a probable SN Ia and an estimate of its redshift, the extraction
of useful cosmological information from SN 1997ff was not feasible and was not
attempted.

\subsection{Serendipity: The NICMOS Campaigns}

Two near-IR assaults on the HDF-N with NICMOS on {\it HST} provided a wealth of
data and understanding on the natural history of galaxies (see Ferguson,
Dickinson, \& Williams 2000 for a review).  The GTO program of Thompson et
al. (1999; GTO 7235) consisted of $\sim$100 orbits of F110W and F160W exposures
of a single $55^{\prime\prime} \times 55^{\prime\prime}$ Camera-3 field,
reaching a limiting AB magnitude (Oke \& Gunn 1983) of 29 in the latter.  The
observations were gathered during 14 consecutive days and the field was
contained within the WF4 portion of HDF-N, {\it serendipitously} imaging the
host of SN 1997ff.  (It is interesting to note that the placement of the GTO
field within the HDF-N had less than a 20\% chance of containing SN 1997ff.)
Although the program did not begin until January 19, 1998, about 25 days after
the discovery of the SN, a series of single-dither exposures (GO 7807) was
taken between the discovery of the SN and the start of the GTO program for the
purpose of verifying the suitability of the chosen guide stars.  Each of these
exposures was for a duration 960~s.  A single F110W and F240M exposure on
Jan. 6, 1998 included the host as did a F160W exposure from Jan. 2, 1998 and
another on Dec. 26, 1997.  {\it The Dec. 26 NICMOS exposure was coincident
within hours of the WFPC2 discovery exposures}.  (It is of further interest to
note the low likelihood of the chance temporal coincidence of the HDF-N SN
Search and the GTO program, each initially scheduled in different {\it HST}
cycles.)

\begin{table}[th] 
\begin{center}
\vspace{0.2cm}
\begin{tabular}{lll}
\multicolumn{3}{c}{Table 1: AB magnitudes of the host galaxy of SN 1997ff} \\
\hline
\hline
bandpass & ${\rm log} (\lambda/ \mu{\rm m})$ & AB mag \\
\hline
$FUV_{160}^b$ & $-$0.80 & $<$30.0 \\
$NUV_{250}^b$ & $-$0.60 & $<$29.2 \\
$U_{300}$ & $-$0.53 & 27.84$^{+2.48}_{-0.70}$ \\
$B_{450}$ & $-$0.34 & 26.67$\pm0.16$ \\
$V_{606}$ & $-$0.22 & 25.64$\pm0.04$ \\
$I_{814}$ & $-$0.10 & 24.42$\pm0.02$ \\
$J_{110}$ & 0.04 & 22.60$\pm0.02$ \\
$J_{125}^a$ & 0.10 & 21.96$\pm0.03$ \\
$H_{160}$ & 0.20 & 21.59$\pm0.01$ \\
$H_{165}^a$ & 0.22 & 21.55$\pm0.03$ \\
$K_s^a$ & 0.33 & 21.03$\pm0.02$ \\

\hline
\hline
\multicolumn{3}{l}{$^a$Ground-based observation; Fernandez-Soto et al. (1999).} 
\\ 
\multicolumn{3}{l}{$^b$95\% limits from {\it HST} STIS; Ferguson (2001, private 
communication).}
\\
\end{tabular}
\end{center}
\end{table}

A second program was undertaken 6 months after the GTO program, between June 14
and June 22, 1998 by Dickinson et al. (2001; GO 7817).  This program observed 
the
entire HDF-N in F110W and F160W to a limiting AB magnitude of $\sim$ 26.5
by mosaicing Camera 3 of NICMOS to study a wider field of galaxies.  This
program also contained the host galaxy and the greatly faded light of the
supernova.  The space-based near-IR photometry of the SN host offered greater
precision and coverage of the SED than the ground-based data alone and allowed
an improved estimate of the photometric redshift.  Using the space-based
$U_{300}B_{450}V_{606}I_{814}J_{110}H_{160}$ photometry and the ground-based
$J_{125}H_{165}K_s$ photometry contained in Table 1, Budav\'ari et al. (2000)
determined the redshift of the host to be $z=1.65\pm0.15$ from fits to either
galaxy SED eigenspectra or these same eigenspectra mildly corrected to improve
the agreement between spectroscopic and photometric redshifts.  The redshift
constraints derived from the host photometry are analyzed in \S 3.1.

\subsection{SN Photometry}

The two HDF-N NICMOS campaigns taken together offer a rare opportunity to
measure the behavior of a supernova at a redshift not accessible from the
ground and perhaps to discriminate between the influence of dark energy and
contaminating astrophysical effects at $z > 1$.  For favored cosmological
models ($\Omega_M \approx 1/3, \Omega_\Lambda \approx 2/3$), a SN~Ia at
$z=1.65$ is expected to peak in F160W at $\sim$24 mag and in F110W at
$\sim$24.5 mag. The 130 kiloseconds of exposure time in each bandpass of the
GTO program would be expected to reach $S/N \approx 100$ for a SN Ia at peak,
though actual measurements impacted by the shot noise of the bright host may be
more uncertain.  Because the SN would not be at peak for some or all of the
observations, we expect further reductions in the measurement precision.  If
the apparent brightnesses of high-redshift SNe are dominated by evolution
and/or dust, and not by cosmology, the $S/N$ of the SN might be further
reduced.  For the single dithers used to test guide stars we expect a $S/N$ no
better than $\sim$10 and possibly worse due to the above mitigating factors.

Another valuable and fortuitous feature of the NICMOS campaigns is that they
likely sample the rest-frame light of the supernova in the $B$ and $V$ bands,
the most studied and best understood wavelength region of nearby SNe.  Indeed,
the great difficulty in observing these wavelengths from the ground has
generally limited the detection and monitoring of SNe Ia to $z \lesssim 1$.  In
the $\sim$2 rest-frame months expected to have elapsed between the two separate
NICMOS programs, a SN Ia is expected to fade $\sim$3 mag, resulting in a
significant surplus of flux in the difference image of the two epochs.

  Our goal is to measure the photometry of the SN throughout the 35 days of the
GTO program.  However, our task is complicated by the proximity of the SN to
its bright host.  GNP99 found that the SN was located at the half-light radius
of the galaxy in F814W.  As we will find, the host contains 2 to 6 times as
much flux at the position of the SN as the peak of the SN PSF, depending on the
band and the date of the exposure.  The strategy of digitally subtracting an
image of the host obtained when the SN has faded (template image) from one
taken when the SN is relatively bright has been successfully employed by the
SCP (Perlmutter et al. 1995) and the High-z Team (Schmidt et al. 1998), as well
as by GNP99 using the original HDF-N images as a template image.  This is the
method we employed.

The task of geometrically mapping (i.e., registering) the images from the
Thompson et al. campaign to align with the template image from the Dickinson et
al. campaign was complicated by the timing and field location of the former.
As seen in Figure 2, the exposure time for the SN in the Thompson et
al. campaign was dispersed irregularly over a 35-day time interval, requiring
careful consideration of the optimal way to measure the temporal behavior of
the SN while still yielding robust photometry (see below).  In addition, the
location of the SN was always extremely close to the corner of the Camera 3
field of $256 \times 256$ pixels, missing the chip during a third of the
dithers, and landing 1 to 15 pixels from the corner in the rest.  Although the
Camera-3 field is remarkably distortion-free in its interior, a mild degree of
``pincushion'' distortion exists in the extreme corners.  Even distortions of
a few tenths of a pixel are intolerable for the accurate subtraction of the
host flux from the SN (Cox et al. 1997).  Although application of the NIC-3
geometric distortion map removes much of the distortion in the field corners,
we applied an empirically derived linear mapping between the SN image and
template to further reduce host contamination.

Our first step was to use the SExtractor algorithm (Bertin \& Arnouts 1996) to
detect sources and measure their centroids in the template and SN images.
Custom software was used to match identical sources in the two lists (Schmidt
et al. 1998).  Next, the {\it geomap} routine in IRAF was used to derive a
flux-conserving mapping of the SN images to the template coordinate system.  In
practice we found that many of the individual 900-s dithers in the GTO campaign
did not provide the desired $S/N$ in the centroid measurements to derive a
robust mapping to the template coordinate system.  A tenable alternative is to
first drizzle together a subset of the dithers (Fruchter \& Hook 1997; Thompson
et al. 1999; GNP99) before deriving the nonlinear coordinate transform.  This
practice has the advantage of increasing the precision of the centroid
measurements of the sources, allowing for the critical sampling of the PSF,
reducing the effects of NICMOS interpixel sensitivity (Storrs et al. 1999), and
providing the ability to further remove cosmic rays (Thompson et al. 1999).
However, the obvious disadvantage of combining the dithers before further
processing (i.e., binning) is a reduction in our ability to resolve the
temporal behavior of the SN.  After much experimentation, we chose an
intermediate strategy of combining the dithers from the main GTO campaign into
3 temporal bins of observed-frame width $\sim 2$ days for each of the F110W and
F160W data sets.  A similar strategy was used by GNP99 to provide time-resolved
magnitudes of the SN in F814W.  The necessary exception to the practice of
binning was for the measurement of the SN flux in the individual 960-s dithers
used to test guide stars before the GTO campaign.

We employed the Alard (2000) algorithm to match the PSF, mean intensity, and
background in the template and SN images. Using the nearest visible sources to
the SN's position, we first derived and then applied a convolution kernel with
a linear variation across the sources. (Again, experimentation showed a
constant convolution kernel was inadequate for matching the two image PSFs, and
second-order convolutions were unstable due to a lack of enough sources with
sufficient $S/N$ to measure the kernel variation.)  Next, we subtracted the
template image from the SN images.  As seen in Figure 3, the resulting image
contains the SN without the contaminating light of the host.  An important test
of these image-processing routines is to verify a lack of significant flux
residuals in the vicinity of other galaxies (which did not host SNe) in the
field.  Confirmation of this test can be seen in Figure 3 (see also GNP99,
Figure 1 for the comparable F814W discovery images of SN 1997ff).

Next, we measured the flux of the SN.  Again, the format of the NICMOS
campaigns (optimized for galaxy studies, not for the monitoring of SN 1997ff)
presented challenges rarely encountered by past high-redshift supernova
programs.  Typically, the leading source of noise in the measurements of the
flux in high-redshift SNe is the shot noise in the sky (or the host galaxy in
{\it HST} observations).  For the NICMOS observations of SN 1997ff, the
dominant source of uncertainty is the host galaxy residuals in the difference
image.  The location of the SN near the core of a much brighter host results in
the appearance of significant galaxy residuals from typically tolerable errors
of 0.1 pixels in the image pair registration.  Eradicating registration errors
at the position of the SN is made more difficult by the location of the host
galaxy near the corner of the Camera 3 field in the Thompson et al. (1999)
campaign.  For some images we reduced this error by processing only a subset of
the image around the SN.  Galaxy residuals in the difference image can also
result from the undersampling of Camera 3 and from variations of the intrapixel
sensitivity.  To derive a robust measure of the SN flux we used a relatively
large aperture to contain the net flux near the SN impacted by the galaxy
residuals.  In practice apertures with a radius of 5 to 10 oversampled-by-two
pixels ($0.5^{\prime\prime}$ to $1.0^{\prime\prime}$) were used.  The flux of
the SN was measured relative to that contained in an equal-sized aperture of a
bright comparison star at the center of the GTO field.  Flux uncertainties were
determined by a Monte Carlo exercise of adding and measuring artificial SNe in
the field with the same brightness and background as SN 1997ff (Schmidt et
al. 1998).  For the single F110W dither from Jan. 6, 1998, the SN was not
detected but a flux upper limit was established by the Monte Carlo exercise.

   We transformed the relative SN flux onto the F110W AB and F160W AB magnitude
systems by applying the zero points of the transformation equation from
Dickinson et al. (2001).  A count rate for the comparison star of
3.22 and 1.85 ADU per second was measured in F110W and F160W, respectively.
Using the zero points of 22.89 for AB F110W and 22.85 for AB F160W gave a
measured mag of 21.62 and 22.18 for the star in these two bandpasses,
respectively.  (An expected uncertainty of $\sim$5\% in these zero points is
insignificant in comparison to the uncertainties in the relative photometry.)
Addition of the measured magnitude differences between the star and the SN
yielded the measurements for the SN on the AB system.  To transform the AB
system magnitudes onto the Vega system we calculated the bandpass-weighted
magnitudes of spectrophotometry of Vega (relative to a flat spectrum) and
derived the zero-point offsets of $-1.34$ and $-0.75$ mag for F160W and
F110W, respectively.  The Vega-system F160W and F110W magnitudes are
given in Table 2 as are the F814W Vega system magnitudes from GNP99.  It is
important to note, that the magnitudes of the SN in F110W and F160W as
listed in Table 2 are {\it underestimates} of the flux due to the presence of SN 
flux
in the template images from Dickinson et al.  The templates were obtained 177
days after the discovery epoch of Dec. 23, 1997.  To accurately correct the SN
mags in Table 2 for the oversubtraction, it is necessary to fit the light curve
to determine this correction.  This step is performed in \S 3.2.

\begin{table}[th]  
\begin{center}
\vspace{0.5cm}
\begin{tabular}{lllllc}
\multicolumn{5}{c}{Table 2: Vega magnitudes of SN 1997ff$^*$} \\
\hline
\hline
Days since 12/23/97 & F814W  & F110W$^*$ & F160W$^*$ & $K_B^d$ & $m_{eff}(B)$ \\
     & (mag) & (mag) & (mag) & (mag) & (mag) \\
\hline
$-$770. &       $>$27.3$^b$ & --- & ---   & --- & --- \nl
$-$229. &       $>$26.0$^a$ & --- & ---  & --- & --- \nl
0.83 &       26.94(0.15)$^c$ & --- & --- & 0.96 & 25.98(0.15)   \nl
2.78 &       27.09(0.15)$^c$ & --- & --- & 1.01 & 26.08(0.15)\nl
3.0  &       --- & --- & 23.49(0.20)    & $-$2.57 & 26.00(0.20)   \nl
3.74 &       26.92(0.13)$^c$ & --- & --- & 1.03 & 25.89(0.13)  \nl
10.  &       --- & --- & 23.56(0.15)  & $-$2.69 & 26.17(0.15)    \nl
14. &     --- & $>$23.3    &  ---     & $-$1.64 & $>$ 24.86\nl
28.5 &      --- & --- &  23.84(0.20)  & $-$3.14 & 26.83(0.20)    \nl
30.5 &      --- & --- &  24.26(0.25)  & $-$3.20 & 27.30(0.25)   \nl
32.0 &     --- & 25.67(0.30)    &  --- & $-$1.65 & 27.16(0.30)     \nl
32.5 &     --- & --- & 24.27(0.20)    & $-$3.26  & 27.36(0.20) \nl
34.5 &     --- & 25.67(0.20)    &  ---  & $-$1.66 & 27.14(0.20)   \nl
37.0 & --- & 25.92(0.20)    &  ---  & $-$1.66 & 27.38(0.20)   \nl

\hline
\hline
\multicolumn{6}{l}{$^*$Underestimates due to flux in template
observed at +177 days.} \\
\multicolumn{6}{l}{$^a$Estimated from {\it HST} archive images (GO 7588).} \\
\multicolumn{6}{l}{$^b$Estimated from initial HDF-N.} \\
\multicolumn{6}{l}{$^c$From GNP99.} \\
\multicolumn{6}{l}{$^d$$K$-corrections to the $B$-band for best fit: $z=1.7$,
SN discovered one week past maximum brightness.} \\
\end{tabular}
\end{center}
\end{table}

\subsection{Host Spectroscopy}   

   On the nights of 21, 26, and 27 Nov. 2000 UT, about 2 hours of optical
spectroscopy were obtained of the SN host using the echelle spectrograph and
imager (ESI) on Keck II under poor conditions (seeing $\sim 1.3''$ and an
airmass of 1.7--2).  During the nights of 23 and 24 Feb. 2001 UT, 3.5 hours of
optical spectroscopy were gathered using the low-resolution imaging
spectrograph (LRIS; Oke et al. 1995) on Keck I in good conditions (seeing $\sim
0.8''$ and an airmass of 1.4).  An additional 2.5 hours of optical spectroscopy
were obtained with ESI on Keck II on the nights of 26 and 27 Feb. 2001 UT.  The
total data set of optical spectroscopy consisted of 8 hours with a wavelength
coverage of 4000~\AA\ to 10000~\AA.  A small amount of continuum flux was
detected in the composite spectrum with evidence of some minor breaks in the
galaxy SED, but there was no evidence of either a strong break in the galaxy
SED or any emission lines.

Observations were made with the near-IR spectrograph NIRSPEC (McLean et
al. 1998) on Keck II on the nights of 16 and 17 March 2001 UT and on 14 April
2001 UT.  On all dates the $0.76''$ slit was oriented at a position angle (PA)
of $184.6^\circ$ to include a galaxy $\sim3''$ to the North.

In the March campaign, the slit was rotated to a PA such that light from both
the host galaxy and the nearby bright galaxy fell onto the slit. The host
galaxy was then positioned on the slit by first obtaining short-integration
images of the field, then moving the telescope to position the nearby galaxy
onto the slit. Spectroscopic observations were obtained by taking four 600-s
integrations. After the first integration, the slit was moved $15''$ in the
spatial direction on the array for the second integration, then moved back to
the original position where the process was repeated.

In the April campaign, the target was acquired by placing a nearby bright star
onto the slit position desired, and offsetting the telescope $\sim 85''$ 
to place the target at the same position on the slit. The conditions
were excellent, with seeing estimated at $0.45''$ FWHM for the duration of the
observation.  Eight exposures of 900 seconds each were obtained; after each
exposure, the offsets were reversed and the alignment of the star on the slit
was checked using images obtained with the IR slit viewing
camera. Judging by these checks, the telescope offsetting and guiding was
accurate to better than 1 pixel ($\sim 0.14$").  For each exposure, the target
was placed on a different position along the $42''$ slit, to reduce detector
systematics and to allow for more accurate sky subtraction.

All the data were reduced following a procedure very similar to that described
in Pettini et al. (2001).  The data from 2001 March and 2001 April were reduced
independently, and then combined with appropriate weighting into the final
co-added 2-D spectrum. The 1-D spectrum was extracted using
an aperture of $1.2''$.
  
The feature which may be {\it very} tentatively identified as the [O~II]
$\lambda$3727 line at $z=1.755$ falls in a relatively rare region that is
unaffected by bright OH lines in the night sky, although there is a strong sky
emission feature at 1.029~$\mu$m, i.e., just to the red of the putative [O~II]
feature.  The excess emission is present in individual subsets of the data, and
becomes more prominent when the 40 minutes of integration time from 2001 March
are combined with the 120 minutes from 2001 April.  The line is well resolved
at a spectral resolution of $\sim 7.5$~\AA, consistent with expectations for
the [O~II] doublet which has a rest-frame separation of $\sim 2$~\AA. (Single
emission lines in the spectrum of the nearby $z=0.556$ galaxy, and most 
but not all sky-subtraction residuals, are significantly narrower.) This
feature is noteworthy and may function as a useful hypothesis to test with
future observations, but at this time its validity is highly uncertain.

\section{Analysis}

\subsection{The Redshift of the SN and its Host}

As discussed in \S 2, fitting the $U_{300}B_{450}V_{606}I_{814}J_{110}H_{160}$
space-based photometry and the $J_{125}H_{165}K_s$ ground-based photometry of
4-403.0 to galaxy SEDs yields a photometric redshift of $z=1.55$ to $1.70$ with
variations depending on whether the fitted model is based on template galaxy
SEDs (Coleman, Wu, \& Weedman 1980, hereafter CWW) or galaxy eigenspectra
(Budav\'ari et al. 2000).  These fits can be seen in Figure 4.  The
well-constrained fits between model and data indicate a far greater degree of
precision in the redshift ($1\sigma \approx 0.02$) than is empirically found by
comparing photometric and spectroscopic redshifts.  We therefore consider the
empirical dispersion from Budav\'ari et al. (2000) as the measure of the
individual uncertainty.

  Using a set of orthogonal eigenspectra derived from the CWW galaxy template
SEDs yields the lowest redshift, $z=1.55\pm0.15$, and empirically the least
precise and robust (Budav\'ari et al. 2000).  Indeed, significant outliers
occasionally result from the application of this method.  A more robust and
precise redshift estimate comes from the fitting of improved eigenspectra.
These are derived from the CWW SED eigenspectra which are first ``repaired''
(see Budav\'ari et al. 2000) as required to improve the agreement between the photometric and
spectroscopic redshifts.  Mild repairing yields $z=1.65\pm0.13$ (model KL2) and
further repairing (model KL5) provides $z=1.70\pm0.10$ (Budav\'ari et
al. 2000).  The biggest advantage of first repairing the eigenspectra is the
suppression of outliers, yielding a more robust estimate.  In
addition, the relatively simple SEDs of early-type galaxies such as
4-403 generally provide more robust photometric redshifts (Budav\'ari et
al. 2000).  For 4-403.0 we will
adopt $z=1.65\pm0.15$ as a measurement which is representative of the
photometric redshift.
 
   An additional and independent pathway to determine the redshift is from the
SN colors.  For the following analysis we will provisionally adopt the
classification from GNP99 of SN 1997ff as a Type Ia supernova based on its red,
elliptical host galaxy.  However, in \S 4.1 we will analyze the degree to which
this classification is merited.

 As seen in Table 2, coincident or near-coincident measurements of SN 1997ff in
different bands provide an observed $I-H$ color of $3.5\pm0.2$ mag and a $J-H$
color of 1.6$\pm0.2$ mag, $30/(1+z)$ days later in the rest frame.  In Figure 5
we plot these measurements as a function of expected colors of SNe Ia over
their temporal evolution at different redshifts (note that the size of the
point scales with the temporal proximity to $B$ maximum).  SNe Ia are bluest
shortly after explosion and become redder with age.  They reach their reddest
color $\sim$25 days after maximum in these bandpasses and return to a modestly
bluer color during the subsequent nebular phase.  As seen in Figure 5, either
of the observed colors of SN 1997ff is redder than a SN Ia at any phase for $z
< 1$.  The $I-H$ color sets a limit of $z > 1.2$ while the $J-H$ color is more
stringent with $z > 1.4$, both at the $>$95\% confidence level.

   However, the constraint obtained from each observed color treated
independently is less restrictive than if we consider their separation in time.
In this case our lower limit on the redshift comes from assuming that the later
color measurement occurs at the reddest phase of the SN and requiring the
earlier color measurement to be consistent with an earlier, bluer phase of the
SN.  In this way we find $z > 1.45$ at the 95\% confidence level.  An upper
limit on the redshift comes from the colors and the observation that the SN is
declining during the five F160W measurements by $\sim 1$ mag in the $\sim$30
observed-frame days.  As seen in Figure 5, the redshift of the SN must be less
than 2.0 to both match the $I-H$ color and be discovered on the decline.  A
redshift in the range of $1.5 < z < 1.8$ is coarsely consistent with the observed colors
and temporal behavior.

   This simple analysis assumes negligible reddening from the host.  While this
assumption is appropriate for most elliptical hosts, we will consider the
effect of reddening explicitly in \S 4.2. Note that Galactic reddening is very
low toward the HDF-N.

  The above exercise cannot readily be performed if we assume the SED of a
common SN II instead of a SN Ia.  Not surprisingly, a comparison of the
observed colors of SN 1997ff to those expected for a blue SN~II (similar to the
well-observed SN 1979C; e.g., Schmidt et al. 1994) yields poor fits to both
color measurements and their time separation.  For any value of the redshift,
the observed $J-H$ color is far bluer than the color of a common SN~II-P or
SN~II-L (for definitions, see Barbon, Ciatti, \& Rosino 1979) at a phase
dictated by the requirement of matching the earlier $I-H$ color.  A common
SN~II would not match both color measurements of SN 1997ff unless it were
observed shortly after explosion and at $z \approx 2$.  However, the observed
decline of SN 1997ff appears inconsistent with the expected rising luminosity
or plateau phase of a SN~II shortly after explosion.  Reddening of the SN would
result in an even greater difference between the data and a common SN~II.
Based on the observed colors and declining luminosity alone, the identification
of SN 1997ff as a normal SN~II is strongly disfavored, a conclusion which is
further discussed in \S 4.2.

   The blue color, bright magnitude, and observed decline of SN 1997ff might
be consistent with some SNe~IIn (Filippenko 1997), which show considerable
heterogeneity in their light curves (Schlegel 1990; Filippenko 2001, private
communication). However, these objects are rare, and they are not found in
old stellar populations (\S 4.2). Similarly, SN 1997ff was unlikely to be
a SN~Ib or SN~Ic, which occur in very young stellar populations, are
generally redder than SN 1997ff, and are rare; see \S 4.2.

   The spectroscopy of the host presented in \S 2.4 provides some evidence
which is consistent with the preceding redshift determinations and is
potentially more precise, but is currently unreliable.  As seen in
Figure 6a, the optical spectrum of the host suggests two minor breaks which
could be identified with the rest-frame breaks at 2640~\AA\ and 2900~\AA\ due to blends of metals (Spinrad et al. 1997).  A simple $\chi^2$ minimization between the optical
spectrum of the host and the same region of the SED of the $z=1.55$ elliptical
LBDS 53w091 (Spinrad et al. 1997) yields a significant minimum bounded by
$z=1.67$ and 1.79 (3$\sigma$ confidence level).  However, this minimum does not
appear robust and we cannot rule out the possibility that other redshift
matches are possible given other models for the host SED.

  The shape of the extracted continuum (Figure 6b) is consistent with the
broad-band photometry from {\it HST} WFPC-2 and NICMOS observations, and is not
inconsistent with the presence of a break at 4000~\AA\ in the galaxy rest frame
(for $z \approx 1.7$), although for a possible redshift of $z \approx 1.8$, the
region longward of the 4000~\AA\ break is beyond the wavelength range covered
by the NIRSPEC spectrum.

  Although the spectroscopic redshift indicators are suggestive of a match with
the photometric indicators, the quality of the spectra is too low and the
identification of spectral features too uncertain to reach a robust
determination of the redshift from the spectroscopy alone.  Therefore in the
following section we derive constraints from the SN without employing the
spectroscopic redshift indications.

\subsection{Probability Density Functions for SN 1997ff}

  The simple method for constraining the redshift described in the previous
section can be refined to make use of all of the SN photometric data
simultaneously.  By varying the parameters needed to empirically fit a SN Ia,
such as the light-curve shape, distance, redshift, and age, we can use the
quality of the fit to determine the probability density function (PDF) of these
parameters.  An additional component of this fitting process is to include the
known correlation between SN~Ia light-curve shapes and their peak luminosities
(Phillips 1993; Phillips et al. 1999; Riess, Press, \& Kirshner 1996;
Perlmutter et al. 1997).  Examples of this fitting process can be seen in
Figure 7 as applied to SN 1997ff.

   In Appendix A we develop a simple formalism for using the observations of
SN 1997ff and any prior information which is appropriate to determine the PDF
of the parameters of luminosity, distance, redshift, and age commonly used to
empirically model SNe Ia.  This method is quite general and its application to
SN Ia photometry is equivalent to the use of a common light-curve fitting
method such as $\Delta m_{15}(B)$ (Phillips et al. 1999; Hamuy et al. 1996),
MLCS (Riess et al. 1996, 1998), or the ``stretch
method'' (Perlmutter et al. 1997), in cases for which the light-curve
information is more constraining than prior information.  The advantage of this
method is its ability to incorporate prior information (e.g., a photometric
redshift) in a statistically sound way which properly assigns weights to the
relative constraints provided by data and priors.

   Due to the presence of SN light in the Dickinson et al. (2001) template
images taken 177 days after the SN discovery, it is necessary to restore the
flux which is necessarily oversubtracted from the F110W and F160W magnitudes
listed in Table 2.  The size of the correction depends on the redshift, age of
discovery, and shape of the SN light curve.  Therefore, we implemented this
correction during the process of fitting the data to parameterized models.  In
the case of the best-fitting light curve, the underestimate of the SN
magnitudes from the GTO campaign is 0.1 to 0.2 mag (depending on the phase), and for the guide
star test exposures (taken when the SN is $\sim$ 1 mag brighter) the correction
is 0.05 mag.
   
   We determined the PDF for SN 1997ff using the methods outlined in
Appendices A and B, the previously described data, and specific priors we
discuss here.

   Riess et al. (1998) found that the observed peak $B$-band luminosities of
SNe~Ia at low redshift ($0.01 < z < 0.1$) and high redshift ($0.3 < z < 1.0$)
are characterized by distribution functions with $\sigma_M \leq 0.25$ mag.  An
even narrower luminosity function of $\sigma_M = 0.17$ mag for SNe Ia was found
by Perlmutter et al. (1999) for a similar set of low-redshift SNe Ia and an
independent set of high-redshift SNe Ia.  Although the peak luminosity function
of very nearby SNe Ia ($z < 0.01$) includes a low-luminosity ``tail'' populated
by so-called SN 1991bg-like SNe~Ia (Filippenko et al. 1992b; Leibundgut et al.
1993; Modjaz et al. 2001), such SNe which are dimmer at peak by $\sim$2 mag are
undetected in high-redshift, magnitude-limited surveys (Li, Filippenko, \&
Riess 2001a).  We defined a normal function prior for the observed peak
luminosity of SNe~Ia with a standard deviation of 0.25 mag.  Assuming that SN
1997ff was drawn from the same population of SNe~Ia that lower-redshift SNe~Ia
have sampled, we expect this prior to be valid.  If, however, the luminosities
of SNe~Ia have significantly evolved by $z \approx 1.7$, then this should be
apparent in the divergence of the redshift-magnitude relation of SNe~Ia from
cosmological models.  We also considered a much less constraining prior of
$\sigma_M = 0.50$ mag. Though this prior underutilizes our empirical knowledge
of SN~Ia luminosities, it does provide a wider latitude to allow the
photometry of SN 1997ff to constrain the fit to its model
light-curve shape.  (Quantitatively, this prior yields similar results
as a perfectly flat luminosity prior.)   It also provides for a possibly larger dispersion in
peak luminosities of SNe~Ia at higher redshifts.

   As provided in \S 3.1, the photometric redshift of the host galaxy from
space-based $U_{300}B_{450}V_{606}I_{814}J_{110}H_{160}$ photometry and the
ground-based $J_{125}H_{165}K_s$ photometry results in the prior constraint
$z=1.65\pm0.15$ (Budav\'ari et al. 2000).  (This constraint is also consistent
with the host spectroscopy as presented in \S 3.1.)  We determined the PDF of
SN 1997ff both with and without this prior photometric constraint.  The latter
approach, while not optimal for determining the best constraints, does allow us
to determine the SN redshift solely from the SN data and test its compatibility
with the photometric and spectroscopic redshift determinations.

   Finally, because no additional information is available to constrain the
remaining two SN~Ia parameters, distance and age at discovery (see Appendix A),
no further knowledge of these variables was included in the priors.

  By marginalizing the 4-D PDF for SN 1997ff over any 3 parameters,
we determined the PDF for the fourth parameter of interest.  The marginal
probability for the redshift, age at discovery, and luminosity are shown in
Figure 8. The marginalized PDF of the redshift is not a simple function, though
it is strongly peaked near $z \approx 1.7$ and is insignificant outside the
range $1.4 < z < 1.95$.  A much lower local maximum is seen at $z=1.55$.  The
redshift measurement of SN 1997ff can be crudely approximated by
$z=1.7^{+0.10}_{-0.15}$.

The consistency of the three redshift indicators, determined independently from
the galaxy colors, the supernova colors, and the host spectroscopy, provides a
powerful and successful crosscheck of our redshift determination. Excluding any
galaxy redshift information has little impact on the marginalized redshift PDF
of the SN because the SN data are significantly more constraining for the
redshift determination. The cause of the difference in measured photometric
redshift precision lies in the difference in the relative homogeneity of galaxy
and SN Ia colors.  For the galaxy photometric determination, the precision of
this method is limited by the variations of galaxy SEDs beyond those which can
even be accounted for from the superposition of eigenspectra. In contrast,
SNe~Ia colors are far more homogeneous and their mild inhomogeneities are well
characterized, leading to more precise constraints on the photometric redshift.

  From the redshift determination we conclude that SN 1997ff is the
highest-redshift SN observed to date (as suspected by GNP99), easily surpassing the two SNe~Ia at $z=1.2$ from
the HZT (Tonry et al. 1999; Coil et al. 2000) and the SCP (Aldering et
al. 1998).  The statistical confidence in this statement is very high.

   Marginalizing the PDF for SN 1997ff over the age-of-discovery parameter
yields the function shown in Figure 8.  We conclude that the SN was discovered
by Gilliland \& Phillips (1998) at an age of a week past $B$-band maximum with
an uncertainty of $\sim$5 days; however, this estimate is highly non-Gaussian,
as seen in Figure 8.  An additional, local maximum in the marginalized
probability is evident at an age of $\sim$15 days after maximum.  The
possibility that the SN was discovered at this later age corresponds to the
same model for which the weaker maximum in the redshift PDF indicated that
$z=1.55$.  This correlation between the redshift and age parameters is a
natural consequence of the reddening of a SN~Ia as it ages, and is shown in
Figure 9. 

   Little additional constraint on the peak-luminosity/light-curve-shape
parameter is gained from the fit, beyond what is provided by the luminosity
function prior as seen in Figure 8.  The $B$-band light curve of a typical
SN~Ia (e.g., Leibundgut 1988; $\Delta m_{15}(B)=1.1$ mag from Hamuy et
al. 1996; $\Delta=0$ mag from Riess et al. 1998) provides an excellent fit to the SN 1997ff data when
the other three parameters are set to their most likely values as seen in the
middle panel of Figure 7.  By relaxing the prior to $\sigma_M=0.50$ mag, we can
better determine the degree to which the SN fit constrains its possible
light-curve shape, and hence its correlated peak luminosity.  As shown in
Figure 8, the fit to the SN 1997ff data disfavors a fast declining, subluminous
SN~Ia light-curve shape.  Quantitatively this result excludes the hypothesis that SN 1997ff
is as much as 0.5 mag subluminous at peak.

   On the bright limb of the peak-luminosity/light-curve-shape relationship,
the constraints are nearly parallel to the luminosity function prior. The
bright limb is defined by spectroscopically peculiar SNe~Ia known as SN
1991T-like events (Filippenko et al. 1992a; Phillips et al. 1992; Li et
al. 1999), which may be overluminous by 0.3--0.6 mag. It is important to note
that SNe~Ia are neither observed nor expected to reach peak magnitudes of more
than $\sim$0.6 mag brighter than the typical (Hamuy et al. 1996; H\"{o}flich \&
Khokhlov 1996).  Therefore, we do not consider light-curve shape models which
are extrapolated beyond the most luminous SNe~Ia observed locally.

   Of critical importance to cosmological hypothesis testing is the
determination of the luminosity distance to such high-redshift SNe~Ia.  Because
of the significant correlation between the distance and photometric redshift
parameters, and the need to use both of these parameters for cosmological
applications, we determined the 2-D PDF for distance and redshift,
simultaneously.  This function is shown in Figure 10 for the different priors
described above.  Although it is preferable to include prior information from
the galaxy photometric redshift estimate and the observed luminosity function
of SNe Ia, the constraints shown in the distance-redshift plane are only
minimally improved with this information and our subsequent conclusions are
insensitive to these priors.

 We also determined the likelihood function for the distance assuming the
tentative spectroscopic redshift.  We find $m-M=42.15\pm0.34$ mag.  This likelihood function is quite Gaussian within the $2\sigma$
boundaries, but flattens beyond for shorter distances (corresponding to an
older discovery age) and steepens beyond for longer distances (corresponding to
a discovery near maximum).  If we assume the tentative spectroscopic redshift,
we find the age of discovery to be $6\pm2$ days after $B$-band maximum
and tighter constraints
on the peak-luminosity/light-curve-shape parameter.  For the latter, 
 we find the
SN to be $0.05 \pm 0.20$ mag fainter at peak than average, which makes it a
highly typical supernova.

\subsection{Cosmological Constraints}

   In Figure 11 we show the redshift and distance data (i.e., the Hubble
diagram) for SNe Ia as presented by the Supernova Cosmology Project (Perlmutter
et al. 1999) and the High-$z$ Supernova Search Team (Riess et al. 1998).  These
data have been binned in redshift to depict the statistical leverage of the SN
Ia sample.  Overplotted are the cosmological models $\Omega_M=0.35,
\Omega_\Lambda=0.65$ (favored), $\Omega_M=0.35, \Omega_\Lambda=0.0$ (open),
$\Omega_M=1.00, \Omega_\Lambda=0.0$ (Einstein--de Sitter), and an astrophysical
model representing a progressive dimming in proportion to redshift due to grey
dust or simple evolution within an open cosmology.  This model is further
described in \S 4.2.  All data and models are plotted as their difference from
an empty Universe ($\Omega_M=0.0,\Omega_\Lambda=0.0$).

   All models are equivalent in the limit of $z=0$.  Differences in the models
are considerable and detectable at $z>0.1$.  Evidence for a significant dark
energy density and current acceleration is provided by the excessive faintness
of the binned data with $0.3 < z < 0.8$ compared to the open model, a net
difference of $\sim$0.25 mag.

   The lack of SNe Ia at an independent redshift interval, beyond $z=1$,
provides only the slimmest of margins for inferring the need for dark energy.
An alternative explanation for the faintness of SNe Ia at $z \approx 0.5$ is a
contaminating astrophysical effect.  Two often-cited candidates for these
effects are SN evolution and grey intergalactic dust.  Although direct tests
for these effects have thus far yielded little evidence to support either
(Riess 2000; Riess et al. 2000), the standard of proof for accepting vacuum
energy (or quintessence) is high.  A more powerful test for any astrophysical
effect which continues to dim SNe at ever greater redshifts is to observe SNe
Ia at $z>1$ (Filippenko \& Riess 2000).  At these redshifts, the Universe was
more compact and familiar gravity would have dominated cosmological repulsion.
The resulting deceleration at these redshifts would be apparent as a {\it
brightening} of SNe Ia relative to a coasting cosmology or to the
aforementioned astrophysicals effects.  The redshift of SN 1997ff is high
enough to probe this earlier epoch and together with the distance measurement
provides the means to discriminate between these hypotheses.

   In Figure 11 we show the constraints derived from SN 1997ff.  In the
redshift-distance plane the principal axes of the error matrix from the
photometric analysis are not quite perpendicular and the confidence contours
are complex.  Because there is only one object available in this highest
redshift interval we prefer to interpret Figure 11 with broad brushstrokes.

   SN 1997ff is brighter by $\sim1.1$ mag (and therefore closer) than expected
for the persistence of a purported source of astrophysical dimming at $z
\approx 0.5$ and beyond.  The statistical confidence of this statement is high
($> 99.99\%$).  {\it This conclusion supports the reality of the measured
acceleration of the Universe from SNe Ia at $z \approx 0.5$ by excluding the
most likely, simple alternatives.}  To avoid this conclusion requires the addition of
an added layer of astrophysical complexity (e.g., intergalactic dust which
dissipates in the interval $0.5 < z < 1.7$, or luminosity evolution which is
suppressed or changes sign in this redshift interval). Other astrophysical
effects such as a change in the SN Ia luminosity distance due to a change in
metallicity with redshift are also disfavored (Shanks et al. 2001).  Systematic
challenges to these conclusions are addressed in \S 4.

   Other cosmological models which predict a relative dimming of SNe Ia at $z >
1$ such as the ``quasi-steady state hypothesis'' appear to be in disagreement
with this observation (Banerjee et al. 2000; Behnke et al. 2001).  Similarly,
models with relatively high vacuum energy and relatively low mass density are
excluded (e.g., $\Omega_\Lambda \approx 1$, $\Omega_M \approx 0$).  If we
assume an approximately flat cosmology as required by observations of the CMB,
and a cosmological constant-like nature for dark energy, the observations of SN
1997ff disfavor $\Omega_\Lambda > 0.85$ or alternately $\Omega_M < 0.15$.

   SN 1997ff also provides an indication that the Universe was decelerating at
the time of the supernova's explosion.  To better understand this likelihood,
in Figure 12 we show the redshift-distance relation of SNe Ia compared to a
family of flat, $\Omega_\Lambda$ cosmologies.  For such cosmologies, the
transition redshift between the accelerating and decelerating epochs occurs at
a redshift of $ \left[ 2\Omega_\Lambda /\Omega_M \right]^{1/3} -1$ (M.  Turner,
2001, private communication).  For increasing values of $\Omega_\Lambda$, the
transition point (i.e., the coasting point) occurs at increasing redshifts.
The highest value of $\Omega_\Lambda$ which is marginally consistent with SN
1997ff is $\Omega_\Lambda$=0.85 (at the $\sim$3$\sigma$ confidence level), for
which the transition redshift occurs at $z=1.25$, significantly below the
redshift of SN 1997ff.  For the Universe to have commenced accelerating before
the explosion of SN 1997ff requires a value of $\Omega_\Lambda > 0.9$, a result
which is highly in conflict with the SN brightness.  We conclude that, within
the framework of these simple but plausible cosmological models, SN 1997ff exploded when the Universe was still decelerating.  Indeed, the increase in the
measured luminosity distance of SNe~Ia between $z \approx 0.5$ and $z \approx
1.7$, a factor of 4.0, is significantly smaller than in most eternally coasting
cosmologies (e.g., $\Omega_M=0, \Omega_\Lambda=0$) and appears to favor the
empirical reality of a net deceleration over this range in redshift.  However,
a rigorous and quantitative test of past deceleration requires a more complete
consideration of the possible nature of dark energy and is beyond the scope of
this paper.

  The above conclusions are unchanged if we adopt the tentative spectroscopic
redshift of the SN host in place of the photometric redshift
indicators (in Figures 11 and 12 the contours would be replaced by a
point at $z=1.755$, $\Delta (m-M)=-0.74\pm0.34$).  In this case the redshift uncertainty is greatly
diminished and the distance uncertainty is mildly reduced.  However, the
dominant source of statistical uncertainty in the testing of cosmological
hypothesis remains the distance uncertainty, not the redshift uncertainty.

   A great deal of theoretical effort has recently been expended to understand
the nature of dark energy.  Some of the possibilities include Einstein's
cosmological constant, a decaying scalar field (``quintessence''; Peebles \&
Ratra 1988; Caldwell et al. 1998), and so on.  A large sample of SNe Ia
distributed over the redshift interval of $0.5 < z < 2.0$ could empirically
break degeneracies between these models (by distinguishing among different
average equations of state, $w = P/\rho c^2$, where $P$ is the pressure and
$\rho$ is the density) if theory alone is insufficient to explain dark energy.
SN 1997ff is in the right redshift range to discriminate between different
dark-energy models, and if one assumes that high-redshift SNe are tracing the
cosmological model and not an astrophysical effect, then SN 1997ff may be
useful for this task.  However, the large uncertainty present in the
measurement of only one SN Ia provides very little leverage to discriminate
between dark-energy models at this time.

\section{Discussion}

The results of \S 3 indicate that SN 1997ff is the most distant SN Ia observed
to date with a redshift of $z=1.7^{+0.1}_{-0.15}$. Moreover, an estimate of its
luminosity distance is consistent with an earlier epoch of deceleration and is
inconsistent with astrophysical challenges (e.g., simple evolution or grey
dust) to the inference of a currently accelerating Universe from SNe Ia at
$z \approx 0.5$.  In this section we explore systematic uncertainties in these
conclusions.

\subsection{SN Classification}

  SNe are generally classified by the presence or absence of characteristic
features in their spectra.  For example, SNe~Ia are distinguished by the
absence of hydrogen lines and the presence of Si~II $\lambda$6150 absorption
(see Filippenko 1997 for a review). Unfortunately, our inability to observe the
defining regions of a SN SED at high redshift necessitates the use of
additional indicators of SN type.

  However, an alternate way to discriminate some SNe Ia is from the morphology
of their host galaxy and its associated star-formation history.  While all
types of SNe have been observed in late-type galaxies, {\it SNe Ia are the only
type to have been observed in early-type galaxies}.  Although this lore is well
known by experienced observers of SNe, this correlation is empirically apparent
from an update of the Asiago SN catalog (Cappellaro et al. 1997; Asiago
website) and can be seen in Figure 13.  Of the $>$1000 SNe whose type and host
galaxy morphologies are all well-defined and have been classified in the modern
scheme, there have been no core-collapse SNe observed in early-type galaxies
(that is, only SNe~Ia have been found in such galaxies).  All $\sim$40 SNe in
elliptical hosts, and classified since the identification of the SN Ia
sub-type, have been SNe Ia.  The same homogeneity of type is true for the
$\sim40$ SNe classified in S0 hosts.  Core-collapse SNe (types II, Ib, and Ic)
first appear along the Hubble sequence in Sa galaxies, and even within these
hosts they form a minority and their relative frequency to SNe Ia is suppressed
by a factor of $\sim 6$ compared to their presence in late-type spirals
(Cappellaro et al. 1997, 1999).  Evolved systems lose their ability to produce
core-collapse SNe.

   The explanation for this well-known observation is deeply rooted in the
nature of supernova progenitors and their ages.  Unlike all other types of SNe
which result from core collapse in massive stars, SNe~Ia are believed to arise
from the thermonuclear disruption of a white dwarf near the Chandrasekhar limit
and thus occur in evolved stellar populations (see Livio 2000 for a
review). The loss of massive stars in elliptical and S0 galaxies, without
comparable replacement, quenches the production of core-collapse SNe, while SNe
Ia, arising from relatively old progenitors, persist.  From the host type of SN
1997ff we might readily conclude, as did GNP99, that it is of Type Ia.

  However, more careful consideration is needed to classify SN 1997ff.  Due to
its high redshift, we need to determine the degree of ongoing star formation
and hence the likelihood of the appearance of a core-collapse SN from a young,
massive star.  While the question of how and when elliptical galaxies form and
evolve is beyond the scope of this paper, we are only concerned with the nature
of the star formation history and stellar populations in the host of SN 1997ff.

   One study of the host galaxy was made previously by Dickinson (1999; see
also Dickinson et al. 2001), who used IR NICMOS photometry to compare the
rest-frame $B-V$ colors of high-redshift HDF-N ellipticals to those expected
for different formation and evolution scenarios.  Dickinson found the
host and another nearby red elliptical to have rest-frame $B-V$ colors
consistent with a burst of star formation at $z \approx 4$ or 5 followed by
passive evolution.

  Here we extend this analysis by comparing the complete
ultraviolet-optical-infrared SED of the host galaxy to Bruzual \& Charlot
(1993) population synthesis models.  The upper panel of Figure 14 superimposes
the galaxy photometry with models that assume a single, short burst of star
formation with a Salpeter initial mass function.  Such a model has negligible
ongoing star formation after the initial burst, and thus its rest-frame
ultraviolet (UV) to optical colors redden as quickly as possible.  After 1~Gyr
has elapsed, this model approximately matches the observed-frame colors of the
host galaxy; a burst occuring 0.5 Gyr or 2.0 Gyr before the SN appears too
short and too long, respectively.  We expect very few remaining massive stars
$\geq 1$~Gyr after the cessation of star formation, and therefore a negligible
chance that SN 1997ff could be a core-collapse SN (whose progenitors live for
less than 40 Myr).  An alternative history would extend the star-formation
timescale, providing a small residual of ongoing star formation to boost the UV
flux while allowing the rest-frame optical colors to redden to match the $IJHK$
photometry.  The bottom panel of Figure 14 shows such a model, with an
exponential star-formation timescale of 0.3 Gyr.  This model matches the
observed SED at an age between 2.0 and 2.5 Gyr (with the far-UV limit favoring
the older age).  Normalized to the $H$-band magnitude of the galaxy, and
assuming $\Omega_M = 0.3$, $\Omega_\Lambda = 0.7$, and $H_0 = 70$ km s$^{-1}$
Mpc$^{-1}$ to compute the luminosity distance, this model provides an ongoing
star formation rate of 0.7 to 0.2~$M_\odot$~yr$^{-1}$ at the time SN 1997ff
exploded [(2.4--10) $\times 10^{-4}$ times the initial rate].  The remaining
population of massive stars should produce 0.004 to 0.001 core-collapse SNe per
year (in the rest frame).  We employ a more empirical route to determine the
expected rate of SNe~Ia due to our inability to identify conclusively their
progenitor systems.  Estimates for the rate of SNe Ia at high redshift from
Pain et al.\ (1996) yield 0.48 SNe Ia per century per $10^{10}$ solar blue
luminosities ($H_0 = 70$~km~s$^{-1}$~Mpc$^{-1}$).  Sullivan et al.\ (2000) and
Kobayashi, Tsujimoto, \& Nomoto (2000) predict a rise in this rate by a factor
of $\sim$2 at the redshift of SN 1997ff.  The host galaxy luminosity is $M_B =
-21.9$, and hence we expect a rate of $\sim 0.07$ SNe Ia per year.  We thus
expect the host to produce 20 to 70 times as many SNe Ia as core-collapse SNe
at the time SN 1997ff exploded (with the far-UV limit favoring the larger
ratio), favoring its classification as a SN Ia independent of the cosmological
model.

  A longer timescale of star formation pushes the time of the initial burst
uncomfortably close to the formation of globular clusters without significantly
altering the expected production ratio of core-collapse SNe to SNe Ia.

  The very tentative identification of a noisy spectral feature with [O~II]
emission would provide an [O~II] flux $f({\mathrm [O~II]}) \approx 8.7\times
10^{-18}$~erg~s$^{-1}$, with a very large uncertainty (at least 50\%) due to
its low $S/N$. As discussed by Kennicutt (1998), the conversion from [O~II]
line flux to star-formation rate is imprecise, and large variations in
[O~II]/H$\alpha$ (as much as 0.5--1 dex) are seen among local galaxies.
Nevertheless, adopting Kennicutt's conversion for a Salpeter initial mass
function (IMF), and assuming (as in \S4.1) a cosmology with $\Omega_M = 0.3$,
$\Omega_\Lambda = 0.7$ and $H_0 = 70$~km~s$^{-1}$~Mpc$^{-1}$, we estimate a
host galaxy star-formation rate of 2.6~$M_\odot$~yr$^{-1}$ from the tentative
[O~II] line identification.  This is 4 to 13 times larger than the rates we
estimated from the broad-band photometric modeling, and would result in a rate
of core-collapse SNe of $\sim 0.01$~yr$^{-1}$, still a factor of $\sim$ 10
smaller than our estimate of the Type Ia SN rate.  However, we consider the
identification of this spectral feature as [O~II] emission very tentative and
the putative flux very uncertain, and therefore the calculation of its implied
star-formation rate is highly speculative.

  Another route to estimating the fraction of core-collapse
progenitors comes from a direct conversion of rest-frame
UV flux (e.g., from the $B_{450}$ magnitude, which corresponds to
roughly 1650~\AA\ in the host-galaxy rest frame at $z=1.7$) to star-formation
rates, following the conversion for a Salpeter IMF from Kennicutt (1998).  This
yields an estimated star-formation rate of $0.8~M_\odot$~yr$^{-1}$, consistent
with the results previously derived using an extended star-formation scenario.
However, this value is likely to be an overestimate, since the standard
conversion factor is estimated from models with constant star-formation rates,
while the galaxy colors resemble those of an early-type galaxy whose current
star-formation rate is almost certainly far lower than its past average.  Some
significant fraction of the UV light may therefore come from longer-lived
stars, and less need be attributed to ongoing star formation.

   A potentially powerful tool to discriminate between SN types 
 comes from enlisting the observed SN data set.  Both the High-$z$ Supernova Search Team and the Supernova
Cosmology Project have relied on the photometric behavior of a SN when a useful
spectrum was not available (Riess et al. 1998; Perlmutter et al. 1999).  The
distance-independent observables of color and light-curve shape have the
potential to discriminate Type Ia SNe from other SN types.  As discussed in \S
3.1, from the observed colors and decline of SN 1997ff we conclude that its
photometric behavior is inconsistent with a Type II supernova at any
redshift. Also, the scarcity of SNe~IIn having similar photometric properties
argues against SN 1997ff being a SN~IIn on photometric grounds. In contrast,
the goodness-of-fit between SN 1997ff and an empirical model of a SN Ia at
$z=1.7$ discovered a week after maximum and with a typical light-curve shape
(see middle panel of Figure 7) is highly consistent with the SN Ia identification
(reduced $\chi^2=0.5$ for $\sim$10 degrees of freedom).  Coupled with the
apparent consistency of the two photometric redshifts and the tentative
spectroscopic redshift, we might consider the SN data to be the best arbiter of
type.  

   Unfortunately, when using only photometric information it may be possible to
confuse a SN Ia with a luminous SN~Ic or SN~Ib (Clocchiatti et al. 2000; Riess
et al. 1998).  However, SNe~Ib and Ic are far rarer than SNe~Ia (Cappellaro et
al. 1997, 1999) and they are expected to arise from even more massive
progenitors than SNe~II (e.g., Wolf-Rayet stars); if so, these progenitors
should be even less populous than those of SNe~II in the comparatively red host
of SN 1997ff. (However, the masses of progenitors of SNe~II and SNe~Ib/Ic can
overlap, if the hydrogen envelope of the progenitor can be lost through mass
transfer in a binary system; e.g., Filippenko 1997)  Empirically, SNe Ib and
Ic are only common in very late-type galaxies (i.e., Sc; Cappellaro et
al. 1997, 1999), environments of marked contrast to the host of SN 1997ff. The
rare, peculiar, highly luminous supernovae (``hypernovae'') that may be
associated with gamma-ray bursts, such as SN 1998bw (e.g., Galama et al. 1998;
Iwamoto et al. 1998; Woosley, Eastman, \& Schmidt 1999), SN 1997cy (Germany et
al. 2000; Turatto et al. 2000), and SN 1999E (Filippenko 2000), also seem to be
produced by core collapse in very massive stars.

   Based on the nature of the host galaxy (an evolved, red elliptical)
and diagnostics available from the
observed colors and temporal behavior of the SN, we find the most likely
interpretation is that SN 1997ff was of Type Ia.

\subsection{Caveats}

  Because we have employed the SN colors to seek constraints on the SN redshift
and age at discovery, both of which are strong functions of SN color, we cannot
employ the colors to determine if there is any reddening by interstellar dust.
The HDF-N was chosen (at high Galactic latitude) in part to minimize
foreground extinction, so we assume that Milky Way reddening of SN 1997ff is
negligible.  Similarly, given the evolved nature of the red, elliptical host,
we assume negligible reddening by the host of the SN.  However, the apparent
consistency with past cosmological deceleration and the apparent inconsistency
with contaminating astrophysical effects reported here would not be challenged
by unexpected, interstellar reddening to SN 1997ff.  To demonstrate this
conclusion, we reddened the SN by $A_B$=0.25 mag in the rest frame and
recalculated the PDF in the distance-redshift plane.  As shown in Figure 11,
the fit is shifted along a ``reddening vector'' further away from the model of
astrophysical effects or cosmological non-deceleration.  While we consider
solutions along the reddening vector less likely, they are important to
bear in mind when assessing quantitative estimates of cosmological parameters
based on the previous analysis.

  Our empirical model of evolution or intergalactic grey extinction (in
magnitudes) is highly simplistic (i.e., linear), and consists only of the
product of a constant and the redshift.  Relative to the empty cosmology
($\Omega_M=0, \Omega_\Lambda=0$), this constant is chosen to be 0.3 mag per unit
redshift to match the observed distances of SNe Ia at $z \approx 0.5$.  The
functional form of this model is the same as that derived by York et al. (2001)
from the consideration of a dust-filled Universe and is shown to be valid for
redshifts near unity.  It also approximates the calculations of Aguirre
(1999a,b).  However, depending on the epoch when the hypothetical dust is
expected to form, the optical depth might be expected to drop at a redshift
higher than two.

  Evolution is far more difficult to model and predict (H\"{o}flich et al.
1998; Umeda et al. 1999a,b; Livio 2000).  For this hypothetical astrophysical
effect our model is that the amount of luminosity evolution would scale with
the mean age available for the growth of the progenitor system (for $z \approx
1$).  While more complex parameterizations are possible, the salient feature of
our simple luminosity evolution model is its monotonic increase with redshift.

    Drell et al. (2000) considered somewhat more complex
phenomenological models of evolution (in magnitudes) consisting of a variable
offset and a variable coefficient multiplied by the logarithm of ($1+z$).  While
the functional form of this model may be less motivated by considerations of
the natural scaling of the physical parameters involved (e.g., time or
metallicity), the additional free parameters make it possible to empirically fit
both the observed dimming ($z \approx 0.5$) and the observed brightening ($z
\approx 1.7$) within a wider range of underlying cosmological models (e.g.,
Einstein--de Sitter).  To test higher-order parameterizations of evolution
than the one we considered here will require measurements of more distant
SNe~Ia in new redshift intervals.

  As discussed in Appendices A and B, constraints derived from the photometry
of SN 1997ff can be recovered from any of the methods used to characterize the
relationship between SN~Ia light curves, color curves, and luminosity.  To test
the sensitivity of our analysis to the light-curve shape method employed, we
rederived the SN constraints in \S 3 using the ``stretch method'' described by
Perlmutter et al. (1997).  The results were in excellent agreement with those
presented using the MLCS method (Riess et al. 1998) in \S 3.  The only
noteworthy exception is that the constraints on the expected luminosity and
distance of SN 1997ff were somewhat narrower for the stretch-method analysis.
The explanation for this difference can be found in the calibration of the
relationship between the peak luminosity and light-curve shape used by each
method.  The stretch method expects a somewhat smaller change in peak
luminosity for a given variation in light-curve shape than the MLCS method.
For the range of possible light-curve shapes allowed by the quality of the fit
to the SN 1997ff data, the stretch method therefore predicts a smaller
variation in luminosity (and hence distance) for SN 1997ff than the MLCS
method.  The cosmological conclusions reported here are supported by either
method.

  Sample selection biases can be important factors to consider when employing
sources detected in a magnitude-limited survey (e.g., Li et al. 2001).
However, because cosmological measurements from SNe~Ia are generally based on
the {\it difference} in the apparent luminosities of SNe from such surveys, a
propagated bias in the cosmological measurements is greatly diminished (Schmidt
et al. 1998; Riess et al. 1998; Perlmutter et al. 1999).  In addition, the
relatively low {\it intrinsic} scatter of SN~Ia distance measurements ($\sigma=0.15$
mag) further reduces such biases.  Monte Carlo simulations of these biases
indicate that the appropriate corrections to the measured distance moduli are 
$< 0.05$ mag, negligible compared to the statistical uncertainties for SN 1997ff presented here.

   It is important to devote special consideration to the likelihood that
SN 1997ff resembles the overluminous and slow declining SN 1991T (Filippenko et
al. 1992a).  An unexpected overluminosity of SN 1997ff would result in
an overestimation of the apparent disagreement with an astrophysical
source of dimming.  The best estimate of the overluminosity of SN
1991T is 0.3$\pm0.3$ mag based on a recent determination of the
Cepheid distance to the host using HST (Saha et al. 2001).  If SN
1997ff were unexpectedly overluminous by this amount, it would still
remain brighter than expected for the dust or simple evolution model
(by about 0.8 mag), but the significance of the difference would be
reduced.  However, the possibility that SN 1997ff matches SN
1991T-like SNe is explicitly included in the analysis in \S 3 by
comparing the fit between the photometry of the former and the latter
(and by employing the widest luminosity prior).  The good fit between
SN 1997ff and a typical SN Ia disfavors its identification
with the slower declining 
SN 1991T.   Empirically, SN 1991T-like events appear to favor
hosts with younger stellar populations (Hamuy et al. 2000; Howell 2000),
but this diagnostic is not as useful as the observed light curve shape
in determining the likelihood that SN 1997ff resembles SN 1991T.

  Clustering of mass in the Universe can cause the line of sight to most SNe to
be underdense relative to the mean, while an occasional supernova may be seen
through an overdense region. In Riess et al. (1998) and Perlmutter et
al. (1999), stochastic lensing which decreased typical fluxes due to underdense
lines of sight to the SNe was considered and found to have little effect on the
cosmological measurements.  The typical deamplification would be larger at $z
\approx 1.7$ and may approach a 10\%--15\% decrease in the observed brightness
of a typical SN such as SN 1997ff (Holz 1998).  For a large sample of SNe, the
mean would provide an unbiased estimate of the unlensed value, but for a single
SN, median statistics are more robust and deamplification of SN 1997ff is
applicable.

   Lensing by the foreground large-scale structure can also alter the apparent
brightness of a distant supernova (Metcalf \& Silk 1999) in the opposite sense
of the preceding consideration. There is a pair of galaxies in the foreground
of SN~1997ff at $z=0.56$, with a separation of $\sim 3''$ and $5.5''$ from the
SN.  If we assume $H_0=70$ km s$^{-1}$ Mpc$^{-1}$, $\Omega_M=0.3$, and
$\Omega_\Lambda=0.7$, these galaxies are found at projected distances of
$\sim20$ and 35 kpc in the lens plane, and have approximately $L^*$ luminosity,
with $M_V \approx -21$, implying a velocity dispersion of $\sim 200$ km
s$^{-1}$.  Assuming that the two galaxies have an approximately isothermal mass
distribution, the resulting magnification of SN~1997ff would be $\sim$0.3 mag,
in good agreement with the results of Lewis \& Ibata (2001).  However, without
detailed knowledge of the form of the mass potential, it is not possible to
recover the precise amplification (and to accurately correct the measured
luminosity).  Our estimate is only an approximation since the mass profile
could fall off more steeply than assumed, or there could be an excess dark
matter concentration associated with this pair of galaxies.

The H$\alpha$ linewidth measured from the NIRSPEC spectrum for the 
nearby $z=0.56$ foreground galaxy is not
significantly resolved at an instrumental resolution of about
180 km~s$^{-1}$ (FWHM), indicating $\sigma \leq 90$  km s$^{-1}$ 
with no indication of any
rotational sheer.  This may not provide a strong constraint on the 
foreground galaxy mass, given uncertainties in the galaxy orientation 
(it appears to be somewhat face--on, although this is difficult to 
assess given its irregular morphology) and the fact that the line 
emission may only trace one star forming region within the galaxy 
rather than the full potential well depth.  We can only say that 
there is no immediate kinematic evidence from existing spectroscopy
for a large mass for the closest foreground galaxy.

   It is possible to derive a useful constraint on the {\it maximum likely}
amplification of the SN by the closest foreground lenses by examining
the shape of the host galaxy which would be
stretched in the tangential direction by an amount that depends on the
SN amplification.  This calculation is performed in Appendix
C and the results can be seen in Figure 15.  From this calculation we conclude
that the {\it lack} of apparent tangential ellipticity of the host galaxy (for
the degree of specific SN amplification estimated above) is not very surprising
($\sim$20\% of randomly selected hosts would exhibit as little tangential
stretching as seen).  However, significantly greater amplifications are not
very likely; 0.6 or 0.8 mag amplifications would produce galaxies with
no
evidence of  tangential stretching only in 6\% and 3\% of identical ensembles,
respectively.  The observed roundness of the image of the host galaxy,
with an axis ratio of 0.85, is further circumstantial evidence against
the presence of substantial lensing.  While unremarkable in
the absence of lensing, such roundness is unusual in images produced
from highly elliptical galaxies oriented so as to counteract the tangential
stretching due to lensing.  We estimate that only $\sim$4\% of
randomly selected hosts will look as round as observed in the presence
of 0.4 magnitudes of amplification, and the fraction drops below 1\%
for a magnification of 0.8 mag.  However, this test is more
circumstantial and therefore less preferable than the tangential
ellipticity test described in Appendix C.

   Together, the specific and stochastic lensing cases result in a possible net
$\sim$0.2 mag shift in the luminosity.  This value is considerably smaller than
the $\sim 1$ mag observed difference between the SN luminosity and that
expected for astrophysical contaminants (grey dust or evolution),
although it is in the direction to reduce this spread.
  Thus, our cosmological conclusions
appear robust to lensing effects, although we cannot rule out 
more exotic lensing scenarios (e.g., a massive dark matter sheet
amplifying the SN but not shearing the host).  Moreover, given that a $\sim$0.2 mag error due to lensing remains a
potential source of systematic uncertainty, we discourage future attempts to
refine the distance measurement to SN 1997ff without a careful consideration of
the full impact of weak lensing.  The challenge posed by
disentangling the affects of lensing and apparent distance provide a
strong impetus to collect more SNe Ia at $z > 1$ to reduce the
statistical impact of such degeneracies.

  We note that the reported consistency with cosmological deceleration and the
exclusion of the dust model or a simple evolution model for SN 1997ff would
remain unchanged in light of a future, high-precision measurement of the
redshift of the host if it is found to be within the redshift interval $1.4 < z
< 1.95$.  This interval is the range over which the SN data are even minimally
consistent with the empirical models of SNe~Ia.  A future redshift
determination which is inconsistent with this range would cast serious doubt on
the interpretation of SN 1997ff as a familiar SN~Ia or any conclusions based on
this interpretation.

  Finally, the detection of SNe~Ia at high redshifts can potentially add
valuable insights into the nature of the progenitors of SNe~Ia. Unfortunately,
with the single case of SN 1997ff, the conclusions are not very restrictive,
and perhaps even circular, since the fact that it occurred in an elliptical
galaxy was used to infer that it is a SN~Ia; many SNe~Ia may be needed,
preferably with spectroscopic confirmation, over a {\it range} of redshifts to
constrain the progenitors of SNe~Ia.  The time allowed for the progenitor
system of SN 1997ff to form and evolve is likely to be more than 1~Gyr (based
on the star-formation history of the elliptical host) and less than about 4 Gyr
(based on the time interval between the redshift of the SN and that of the
initial formation of stars). However, we already know that the progenitors of
{\it some} SNe~Ia are very old ($\sim 10^{10}$ yr), since they are found in
present-day, gas-deficient, old elliptical galaxies. Also, it already appears
that the progenitors of {\it most} SNe~Ia are relatively young ($\lesssim
1$~Gyr), since they are found with greatest frequency in late-type galaxies
(Cappellaro et al.  1997, 1999) and might even exhibit a loose correlation with
spiral arms (Bartunov, Tsvetkov, \& Filimonova 1994).

\section {Conclusions}

  1. SN 1997ff is the highest-redshift SN~Ia observed to date, and we estimate
its redshift to be $\sim 1.7$.  This redshift is consistent with the
measurements made from either the SN data, the 9-band photometric redshift of
the host, or the tentative indication from the host-galaxy spectroscopy.  The
classification as a SN~Ia is derived from observational and theoretical
evidence that the evolved, elliptical host is deficient in the progenitors of
core-collapse SNe.  The classification is also supported by
diagnostics available from the observed colors
and temporal behavior of the SN.

  2. The derived constraints for the redshift and distance of SN 1997ff are
consistent with the early decelerating phase of a currently accelerating
Universe and thus are a valuable test of a Universe with dark energy.  The
results are inconsistent with simple evolution or grey dust, the two most
favored astrophysical effects which could mimic previous evidence for an
accelerating Universe from SNe~Ia at $z \approx 0.5$.

  3. We consider several sources of potential systematic error including
gravitational lensing, supernova misclassification, sample selection bias, and
luminosity calibration errors.  Currently, none of these effects alone appears
likely to challenge our conclusions.  However, observations of more
SNe Ia at $z>1$ are needed to test more complex challenges to the
accelerating Universe hypothesis and to probe the nature of dark energy.

\bigskip
\bigskip 

   We wish to thank Chris Fassnacht, Kailash Sahu, Nick Suntzeff, Bruno
Leibundgut, Steve Beckwith, Sean Carroll, Harry Ferguson, Greg Aldering, and Casey Papovich
for valuable contributions and discussions.  We are indebted to Michael Turner
for his efforts to refine our consideration of a past deceleration. Parts of
this project were supported by grants GO-6473, GO-7817, and AR-7984 from the
Space Telescope Science Institute, which is operated by the Association of
Universities for Research in Astronomy, Inc., under NASA contract NAS~5-26555;
we also acknowledge NASA support of the NICMOS GTO team. A.V.F. is grateful for
a Guggenheim Foundation Fellowship and for NSF grant AST-9987438.  This work was supported by a NASA LTSA grant to PEN and by the Director,
Office of Science under U.S. Department of Energy Contract No.
DE-AC03-76SF00098. PEN also thanks NERSC for a generous allocation of
computer time.

\appendix

\centerline {APPENDICES}

\section {SN and Cosmological Likelihood Function}

   Our goal is to define a simple formalism to combine the popular light-curve
fitting methods with prior information (e.g., a likelihood function) in a
statistically sound way.  The motivation for this method is for application to
sparse or inhomogeneous SN Ia data sets which can be augmented with additional
information.  Current light-curve fitting methods can include four free
parameters (age, distance, extinction, and luminosity), and could include more
such as a photometric redshift or an extinction law which can be constrained
from SN colors.  Alternatively, it may be desirable to discard the two 
parameters
of distance and redshift in exchange for cosmological parameters such as $H_0$,
$\Omega_m$, $\Omega_\Lambda$, $w$, etc.  However, data sets gathered at high
redshift are often noisy, sparse, and inhomogeneous, and may provide a high or
low degree of leverage on any of these parameters.  It may be desirous to
supplement the data set with prior information such as a host galaxy
photometric redshift or a luminosity function, while adding no additional
information to some parameters such as the age at discovery.

   Of course, a powerful language for such formalism is Bayes's theorem.  In
our case,

\bq p({\bf q} | {\bf D}) = {p({\bf D} | {\bf q}) p({\bf q})
 \over p({\bf D})},
 \eq where {\bf D} is a set of SN data and priors, and {\bf q} is a combination
 of SN or cosmological parameters we are trying to constrain.  Because we have
 no prior constraints on {\bf D}, we have

\bq p({\bf q} | {\bf D}) \propto p({\bf D} | {\bf q})
p({\bf q}).\eq
In principle, ${\bf q}$ should contain the set of all SN or cosmological
parameters which cause our data to appear as they do, but in practice it is
only necessary to include parameters to which our data set is sensitive.  Here
we consider the set of parameters $\mu_B$ (apparent distance modulus), $t_d$
(days past $B$-band maximum at the time of discovery), $\Delta$ (the MLCS
luminosity/light-curve-shape parameter; Riess et al. 1996), and
$z$ (the redshift).  Also, {\bf D} is the set ${\bf m}$ of magnitude
measurements in Table 1 for SN 1997ff.

As a result we have 
\bq p(\mu_B,t_d,\Delta,z | {\bf m}) \propto \left(\prod_i
\frac{1}{\sqrt{2 \pi \sigma_{m,i}^2}} \right) \exp \left( - \frac{\chi^2}{2}
\right) \left [ p(t_d)p(\Delta|t_d)p(\mu_B|t_d,\Delta)p(z|\mu_B,t_d,\Delta)
\right] \eq and $\chi^2$ is defined as \bq \chi^2(\mu_B,t_d,\Delta,z)=\sum_i {
(\mu_B+M_B(\Delta,t_d,t_i,z)-m_i+K_i(z,t_i,t_d,\Delta))^2 \over
\sigma_{m_i}^2}. \eq 
The term $M_B(\Delta,t_d,t_i,z)$ is an empirical light-curve 
model of a SN Ia in the $B$ band.  To compare an apparent magnitude
measurement $m_i$ at time $t_i$ we compute the expected magnitude of a SN MLCS
model ${\bf M_B(\Delta)}$ at the rest-frame age relative to $B$-band maximum of
$(t_i - t_d)/(1+z)$.  The term $K_i(z,t_i,t_d,\Delta)$ is a cross-filter
$K$-correction and is discussed extensively for SNe Ia elsewhere (Kim, Goobar,
\& Perlmutter 1996; Riess et al. 1998; Schmidt et al. 1998; Nugent, Kim, \& 
Perlmutter 2001).  It is determined by synthetically calculating the pseudocolor
$X-B$ (where $X$ represents the passband F814W, F110W, or F160W blueshifted by
$1+z$) using the SED of a SN Ia (with intrinsic colors appropriate to a SN Ia
with a luminosity parameter of $\Delta$) at an age relative to $B$ 
maximum of $(t_i - t_d)/(1+z)$.

   The PDFs in brackets on the right-hand side of equation (A3) are priors in
which we can include any desired amount of prior information about these
parameters.  $p(t_d)$ is the prior distribution on the age at which the SN was
discovered given no other information.  A spectrum of a SN Ia can be used to
set a prior constraint on the age (Riess et al. 1997), or without such
information we would choose this prior to be constant.  $p(\Delta|t_d)$ is the
observed luminosity function of SNe Ia.  $p(\mu_B|t_d,\Delta)$ is given by the
area of the hypershell with a radius of $\mu_B$.  In principle this prior,
derived from the area of hypershells, is cosmology-dependent because the
geometry of these hypershells is determined by the cosmological parameters.
However, for SN data which are moderately constraining (such as is the case for
SN 1997ff), the specific cosmology chosen has a negligible impact on the fits.
If this were not the case, it would be necessary to substitute the ``nuisance
parameters'' of distance and redshift for cosmological parameters and the
additional constraint in the form of the equation of the luminosity distance
(Schmidt et al. 1998).  For now treating $z$ and distance as independent, the
term $p(z|\mu_B,t_d,\Delta)$ contains the prior constraint on the redshift
(such as a photometric redshift) and possibly a time dilation effect such as
the $1+z$ expansion in time intervals (and hence the likelihood of finding a SN
Ia at higher redshifts).

   Either the $\chi^2$ statistic or the {\it a posteriori} probability may be
used in the usual ways to determine constraints on any combinations of the
parameters {\bf q} (Press et al. 1992).

\section {Cross-Band $K$-Corrections}

For a full description of the \kcorrs\ and the uncertainties involved with
their application see Nugent et al. (2001). In what follows we briefly
summarize the relevant issues which concern us here.

 The magnitude of a SN~Ia in filter $y$ can be expressed as
the sum of its absolute magnitude $M_x$, cross-filter \kcorr\
$K_{xy}$, distance modulus $\mu$, and extinction due to dust in both
the host galaxy, $A_x$, and our galaxy, $A_y$:
\begin{eqnarray}
m_y(t(1+z)) = M_x(t,s) + K_{xy}(z,t,s)\nonumber \\
    + \mu(z,\Omega_M,\Omega_\Lambda,H_o)+A_x+A_y.
\label{kcorr}
\end{eqnarray}
Here $t$ refers to the epoch when the SN~Ia is being observed and $z$ is its
redshift.  The parameter, $s$, can be any parameter (or method) which is used
to characterize the relationship between the light curves, colors, and
luminosity of an individual SN Ia. For the MLCS method (Riess et al. 1996, 
1998) this parameter would be the $\Delta$
parameter, formally the peak absolute luminosity of the visual light curve
(relative to a fiducial SN Ia).  Alternatively it could be the
$\Delta m_{15}(B)$ parameter (Phillips 1993; Phillips et al. 1999) or the
stretch-factor (as described in Perlmutter et al. 1997).

The spectral template used to produced the \kcorrs\ was created by gathering
together all the spectra of well-observed SNe~Ia currently available to the
authors. Especially important to these calculations were the SNe observed in
the UV by {\it IUE} and {\it HST} (Cappellaro et al. 1995; Kirshner et
al. 1993). A set of standard photometric templates was then created for a
fiducial SN~Ia as well as SNe Ia with values of $s$ chosen to sample and span
the observed range of SNe Ia in $UBVRI$.  The spectra were then ``flux
calibrated'' in order to reproduce the observed magnitudes of these template
light curves by both adjusting the zero point of the flux scale and applying a
slope correction to the flux so that each spectrum would have the correct color
(Riess et al. 1996, 1998) ) for a particular phase. The slope correction was
performed by altering the flux using the reddening law of Cardelli, Clayton, \&
Mathis (1989) (either making them bluer or redder accordingly). For the list of
the SNe~Ia and their corresponding $UBVRI$ magnitudes see Nugent et al. (2001).

To determine the $K$-correction for a given value of $s$ and $z$, the set of
spectra were first flux calibrated (as described above) to match the fluxes of
the set of template light curves described by the parameter $s$.  The
cross-band $K$-correction was then calculated according to the formulae of Kim,
Goobar, \& Perlmutter (1996; see also Schmidt et al. 1998).  After
identification of the observed bandpass and a rest-frame bandpass, a pseudocolor
is numerically calculated.  This pseudocolor is the difference in the magnitude
of the SN in the rest-frame bandpass and the observed-frame bandpass, the latter
blueshifted by ($1+z$), and is calculated from the template SN 
spectrophotometry.
These calculations are derived from spectrophotometry of SNe Ia at different
phases and we interpolate these values to determine the correction at a given
phase.  Examples of these calculations are given in Table 2; they are uncertain
to less than 10\%.

\section {The Tangential Ellipticity Test for the SN Host Galaxy}

  Significant gravitational lensing due to galaxies close to the line of sight
would be expected to cause a detectable distortion --- primarily a tangential
stretch --- of the shape of the host galaxy.  The observed tangential
ellipticity can in turn be used to constrain the amplification due to lensing.
However, the intrinsic shape of the host galaxy is not known, and thus it is
possible that the galaxy is intrinsically elongated in the radial direction, so
that the final image does appear nearly round.

  We can, however, apply a statistical test to determine the likelihood that we
would be able to detect the tangential stretch.  We ask how frequently, for a
randomized distribution of the intrinsic (i.e., unlensed) parameters of the
galaxy, the lensed image will appear to have as little tangential stretch as is
observed.  (Note that this test is one-sided, since we know that the lens causes a
{\it tangential} stretch.)

  For simplicity, we first limit our consideration to lensing from the nearest
bright galaxy, about $3''$ N and $0.6''$ E of the host galaxy, and we assume a
circularly symmetric, singular isothermal mass distribution.  For a mass
distribution of this form, images are not stretched radially, but only
tangentially, and the amplification factor ($\mu$) of the amplification is identical to
the factor by which the tangential component of the host galaxy is stretched.

We define the {\it tangential ellipticity} of a galaxy image as the quantity

$$ \epsilon_{T} = {M_{TT}-M_{RR} \over M_{TT} + M_{RR}} = {1-(b/a)^2
\over 1+(b/a)^2} \cos(2\,\theta), $$ 

\noindent
where $M_{TT}$ and $M_{RR}$ are the second-order moments of the light
distribution in the tangential and radial directions, respectively, $ b/a
$ is the ratio of minor to major axis of the light distribution, and $
\theta $ is the angle between the major axis of the host galaxy and the
tangential direction, $0^\circ \le \theta \le 90^\circ$.  The angle $\theta =
0^\circ$ if the source is tangentially oriented, and $\theta = 90^\circ$ if the
image is radially oriented.  We use a suffix $I$ when referring to
intrinsic quantities, without the distortion due to lensing, and $M$
when referring to measured quantities. 

In the presence of amplification $ \mu > 1 $, the source is stretched
tangentially by a factor $ \mu $, while its radial size remains
unchanged.  The measured ellipticity $ \epsilon_{T,M} $ is then related
to the intrinsic ellipticity $ \epsilon_{T,I} $ as

$$ \epsilon_{T,M} = {\mu^2(1+\epsilon_{T,I}) - (1-\epsilon_{T,I}) \over 
   (1+\epsilon_{T,I}) + \mu^2 (1-\epsilon_{T,I})}. $$

\noindent
The observed tangential ellipticity $ \epsilon_{T,O} $ of the host
galaxy can be estimated from its axis ratio $ b/a = 0.851 $ and position
angle of the major axis, $139.6^\circ$ East of North (Dickinson 2001, private
communication).  Since the lens is at a position angle of $11.3^\circ \pm
0.7^\circ$ from the source, the major axis is $38.3^\circ$ from the
tangential direction, and the observed tangential ellipticity 
$\epsilon_{T,O} = 0.037$ (i.e., quite small; see Figure 1). 

  In order to determine how often the measured ellipticity $ \epsilon_{T,M} $
of a galaxy like the host galaxy will be as small as the observed ellipticity $
\epsilon_{T,O} $, we need to construct an appropriate distribution of the two
relevant intrinsic image parameters, the intrinsic axis ratio $ (b/a)_I $ and
the intrinsic orientation angle $ \theta_I $.  Since there should be no
physical interaction between lens and host galaxy, the orientation $ \theta_I $
is distributed randomly and uniformly in the allowed range $0^\circ \le \theta_I 
\le
90^\circ$.  We draw the axis ratio $(b/a)_I$ from the distribution observed by
Lambas, Maddox, \& Loveday (1992) for galaxies classified as ellipticals in the
APM survey.  Although this distribution is obtained for low-redshift objects,
we expect it to be representative of the high-redshift host galaxy because the
latter is observed in the near-IR (rest-frame optical) and does not appear
significantly perturbed.

  Figure 15 shows a subset ($10^4$ of the $10^6$ samples) of randomly selected
combinations of the intrinsic axis ratio and orientation angle of simulated
host galaxies.  For several different hypothetical values of the SN
amplification, a curve is shown indicating the combinations of the intrinsic axis
ratio and orientation angle resulting in the observed tangential eccentricity
of the SN host.  Intrinsic values above each curve correspond to galaxies whose
elongation and orientation are such as to produce a tangential ellipticity
smaller than the observed value of 0.037.  Since many ellipticals are
intrinsically nearly round ($ b/a > 0.7 $), for large assumed amplifications
the intrinsic position angle must be very close to radial ($\theta_I >
70^\circ$) if the observed ellipticity is to be as small as observed.  Each
curve is labelled with the assumed amplification (in magnitudes) of the SN and
the fraction of randomly selected galaxies that fall above it, i.e., whose
observed tangential ellipticity would be 0.037 or smaller.

  The small degree of observed tangential ellipticity of the SN host provides a
useful constraint for the maximum likely amplification of the SN.  (This
determination is independent of the estimate in \S 4.2 of the {\it expected}
amount of net amplification of the SN by both stochastic and specific lensing,
$\sim$0.2 mag.)  For example, if we assumed that the SN was amplified by 0.6
mag by the foreground lens, we would conclude that we were relatively lucky
(i.e., a 6\% chance) to see such a small apparent tangential ellipticity for
the host (a result which can occur for an E6 or E7 host galaxy intrinsically
aligned within $20^\circ$ of the radial orientation). The chance of seeing such 
a
small apparent tangential ellipticity for an amplification which moves the SN
from the dust-filled Universe model to its observed brightness ($\sim$1.1 mag)
is less than 1\%.
   
  Interestingly, the calculations above have predictive power for the
  other close lens candidate, located $5.3''$ from the host and also
  at $z=0.56$.  This galaxy lies only $20^\circ$ from
the radial axis joining the host and the closest lens.  This lens would distort the host
galaxy in a direction very similar to the one resulting from the closest lens.

\vfill \eject
 
\centerline {\bf References}
\vskip 12 pt

\refitem Aguirre, A. 1999a, ApJ, 512, 19
 
\refitem Aguirre, A. 1999b, ApJ, 525, 583

\refitem Alard, C. 2000, A\&AS, 114,363

\refitem Aldering, G., et al. 1998, IAU Circ. 7046

\refitem Balbi, A., et al. 2000, ApJ, 545, L1 

\refitem Banerjee, S. K., Narlikar, J. V., Wickramasinghe, N. C., Hoyle, F., \&
Burbidge, G. 2000, AJ, 119, 2583

\refitem Barbon, R., Ciatti, F., \& Rosino, L. 1979, A\&A, 72, 287

\refitem Bartunov, O. S., Tsvetkov, D. Yu., \& Filimonova, I. V. 1994, PASP,
  106, 1276

\refitem Behnke, D., et al. 2001,  arXive: gr-qc/0102039

\refitem Bertin, E., \& Arnouts, S. 1996, A\&AS, 117, 393

\refitem Branch, D., Romanishin, W., \& Baron, E. 1996, ApJ, 465, 73 (Erratum:
  467, 473)

\refitem Bruzual, A. G., \& Charlot, S. 1993, ApJ, 405, 538

\refitem Budav\'ari, T., Szalay, A. S., Connolly, A. J., Csabai, I., \&
Dickinson, M. 2000, AJ, 120, 1588

\refitem Caldwell, R. R., Dav\'e, R., \& Steinhardt, P. J. 1998,
Ap\&SS, 261, 303

\refitem Cappellaro, E., Turatto, M., \& Fernley, J. 1995, IUE -- ULDA Access
Guide No. 6: Supernovae (The Netherlands: ESA)

\refitem Cappellaro, E., Evans, R., \& Turatto, M. 1999, A\&A, 351, 459

\refitem Cappellaro, E., et al. 1997, A\&A, 322,431

\refitem Cardelli, J.~A., Clayton, G.~C., \& Mathis, J.~S. 1989, ApJ, 345, 245

\refitem Clocchiatti, A., et al. 2000, ApJ, 529, 661

\refitem Cohen, J. G., et al. 2000, ApJ, 538, 29

\refitem Coil, A., et al. 2000, ApJ, 544, L111

\refitem Coleman, G. D., Wu, C.-C., \& Weedman, D. W. 1980, ApJS, 43, 393 (CWW)

\refitem Cox, C., Ritchie, C., Bergeron, E., MacKenty, J., \& Noll, K. 1997, 
   NICMOS STScI Instrument Science Report 97-07 (Baltimore: STScI)

\refitem Curtis, D., et al. 2000,
Supernova/Acceleration Probe (SNAP), proposal to DOE and NSF,
http://snap.lbl.gov/

\refitem de Bernardis, P., et al. 2000, Nature, 404, 955

\refitem Dickinson, M. 1999, in {\it After the Dark Ages: When Galaxies were
Young (the Universe at 2 $<$ z $<$ 5)}, ed. S. Holt \& E. Smith (New York:
AIP), 122

\refitem Dickinson, M., et al. 2001, in preparation

\refitem Drell, P. S., Loredo, T. J., \& Wasserman, I. 2000, ApJ, 530, 593

\refitem Ferguson, H. C., Dickinson, M., \& Williams, R. 2000, ARA\&A, 38, 667

\refitem Fernandez-Soto, A., Lanzetta, K. M., \& Yahil, A. 1999, ApJ, 513, 34

\refitem Filippenko, A. V. 1997, ARA\&A, 35, 309

\refitem Filippenko, A. V. 2000, in Cosmic Explosions, ed. S. S. Holt \&
  W. W. Zhang (New York: AIP), 123

\refitem Filippenko, A. V., \& Riess, A. G. 2000, in {\it Particle
Physics and Cosmology: Second Tropical Workshop}, ed. J. F. Nieves
(New York: AIP), 227

\refitem Filippenko, A. V., et al. 1992a, ApJ, 384, L15

\refitem Filippenko, A. V., et al. 1992b, AJ, 104, 1543

\refitem Fruchter, A. S., \& Hook, R. N. 1997, Proc. SPIE, 3164, 120

\refitem Galama, T. J., et al. 1998, Nature, 395, 670

\refitem Germany, L. M., Reiss, D. J., Sadler, E. M., \& Schmidt, B. P.
   2000, ApJ, 533, 320
 
\refitem Gilliland, R. L., Nugent, P. E., \& Phillips,  M. M. 1999,
ApJ, 521, 30 (GNP99)

\refitem Gilliland, R. L., \& Phillips, M. M. 1998, IAU Circ. 6810

\refitem Hamuy, M., et al. 1996, AJ, 112, 2408
 
\refitem H\"{o}flich, P., \& Khokhlov, A. 1996, ApJ, 457, 500

\refitem H\"{o}flich, P., Nomoto, K., Umeda, H., \& Wheeler, J. C.
2000, ApJ, 528, 590

\refitem H\"{o}flich, P., Wheeler, J. C., \& Thielemann, F. K. 1998,
ApJ, 495, 617

\refitem Holz, D. E. 1998, ApJ, 506, 1

\refitem Iwamoto, K., et al. 1998, Nature, 395, 672

\refitem Jaffe, A., et al. 2001, astro-ph/0007333

\refitem Kennicutt, R. C. 1998, ARA\&A, 36, 189

\refitem Kim, A., Goobar, A., \& Perlmutter, S. 1996, PASP, 108, 190

\refitem Kirshner, R., et al. 1993, ApJ, 415, 589

\refitem Kobayashi, C., Tsujimoto, T., \& Nomoto, K. 2000, ApJ, 539, 26

\refitem Lambas, D. G., Maddox, S. J., \&  Loveday, J. 1992, MNRAS, 258, 404

\refitem Leibundgut, B. 1988, PhD thesis, University of Basel

\refitem Leibundgut, B., et al. 1993, AJ, 105, 301


\refitem Lewis, G. F., \& Ibata, R. A. 2001, MNRAS, submitted
  (astro-ph/0104254)

\refitem Li, W., Filippenko, A. V., \& Riess, A. G. 2001a, ApJ, 546, 719

\refitem Li, W. D., et al. 1999, AJ, 117, 2709

\refitem Livio, M. 2000, in {\it Type Ia Supernovae: Theory and Cosmology},
eds. J. C. Niemeyer \& J. W. Truran (Cambridge: Cambridge Univ. Press), 33


\refitem McLean, I., et al. 1998, Proc. SPIE, 3354, 566

\refitem Metcalf, R. B., \& Silk, J. 1999, ApJ, 519, 1

\refitem Modjaz, M., Li, W., Filippenko, A. V., King, J. Y., Leonard, D. C., Matheson, T., Treffers, R. R., \& Riess, A. G. 2001, PASP, 113, 308

\refitem Nomoto, K., Umeda, H., Hachisu, I., Kato, M., Kobayashi, C., \&
Tsujimoto, T. 2000, in {\it Type Ia Supernovae: Theory and Cosmology},
eds. J. C. Niemeyer \& J. W. Truran (Cambridge: Cambridge Univ. Press), 63

\refitem Nugent, P. 2000, in {\it Particle Physics and Cosmology:
Second Tropical Workshop}, ed. J. F. Nieves (New York: AIP), 263

\refitem Nugent, P. E., Kim, A., \& Perlmutter, S. 2001, in preparation

\refitem Oke, J. B., \& Gunn, J. E. 1983, ApJ, 266, 713

\refitem Oke, J. B., et al. 1995, PASP, 107, 375

\refitem Pain, R., et al. 1996, ApJ, 473, 356

\refitem Peebles, P. J. E., \& Ratra, B. 1988, ApJ, 325, L17 

\refitem Perlmutter, S., et al. 1995, ApJ, 440, 41

\refitem Perlmutter, S., et al. 1997, ApJ, 483, 565
 
\refitem Perlmutter, S., et al. 1999, ApJ, 517, 565

\refitem Pettini, M., Shapley, A. E., Steidel, C. C., Cuby, J.-G., 
  Dickinson, M., Moorwood, A. F. M., Adelberger, K. L., \& Giavalisco, 
  M. 2001, ApJ, 554, in press (astro-ph/0102456)

\refitem Phillips, M.~M. 1993, ApJ, 413, L105

\refitem Phillips, M. M., Lira, P., Suntzeff, N. B., Schommer, R. A., Hamuy,
  M., \& Maza, J. 1999, AJ, 118, 1766

\refitem Phillips, M. M., et al. 1992, AJ, 103, 1632

\refitem Pinto, P. A., \& Eastman, R. G. 2000, ApJ, 530, 744

\refitem Press, W. H., Teukolsky, S. A., Vetterling, W. T., \&
  Flannery, B. P. 1992, Numerical Recipes in FORTRAN (2nd ed.;
  Cambridge: Cambridge Univ. Press)

\refitem Rana, N. C. 1979, Ap\&SS, 66, 173

\refitem Rana, N. C. 1980, Ap\&SS, 71, 123

\refitem Riess, A. G. 2000, PASP, 112, 1284

\refitem Riess, A. G., Press, W.H., \& Kirshner, R. P. 1996, ApJ, 473,
88 
 
\refitem Riess, A. G., et al. 1997, AJ, 114, 722

\refitem Riess, A. G., et al. 1998, AJ, 116, 1009

\refitem Riess, A. G., et al. 2000, ApJ, 536, 62

\refitem Ruiz-Lapuente, P., \& Canal, R. 1998, ApJ, 497, 57

\refitem Saha, A., et al. 2001, ApJ, in press

\refitem Sandage, A., \& Hardy, E. 1973, ApJ, 183, 743

\refitem Sawicki, M., Lin H., \& Yee, H. 1997, AJ, 113, 1

\refitem Schlegel, E. M. 1990, MNRAS, 244, 269

\refitem Schmidt, B. P., et al. 1994, ApJ, 432, 42

\refitem Schmidt, B. P., et al. 1998, ApJ, 507, 46

\refitem Shanks, T., Allen, P. D., Hoyle, F., \& Tanvir, N. R.
   2001, in New Cosmological Data and the Value of the Fundamental 
   Parameters, ed. A. N. Lasenby \& A. Wilkinson (San Francisco:
   ASP), in press (astro-ph/0102450)

\refitem Spinrad, H., Dey, A., Stern, D., Dunlop, J., Peacock, J., 
  Jimenez, R., \& Windhorst, R. 1997, ApJ, 484, 581

\refitem Storrs, A., Hook, R., Stiavelli, M., Hanley, C., \& 
  Freudling, W. 1999. NICMOS STScI Instrument Science Report 99-05
  (Baltimore: STScI)

\refitem Sullivan, M., Ellis, R., Nugent, P., Smail, I., \& Madau, P. 2000,
  MNRAS, 319, 549

\refitem Thompson, R. I., Storrie-Lombardi, L. J., Weymann, R. J., Rieke,
  M. J., Schneider, G., Stobie, E., \& Lytle, D. 1999, AJ, 117, 17

\refitem Tonry, J., et al. 1999, IAU Circ. 7312

\refitem Turatto, M., et al. 2000, ApJ, 534, L57

\refitem Umeda, H., Nomoto, K., Kobayashi, C., Hachisu, I., \&
  Kato, M. 1999a, ApJ, 522, L43

\refitem Umeda, H., Nomoto, K., Yamaoka, H., \& Wanajo, S.
  1999b, AJ, 513, 861

\refitem Williams, R. E., et al. 1996, AJ, 112, 1335

\refitem Woosley, S. E., Eastman, R. G., \& Schmidt, B. P. 1999, 
  ApJ, 516, 788

\refitem York, T., et al. 2001, in preparation

\refitem Yungelson, L. R., \& Livio, M. 2000, ApJ, 528, 108

\vfill \eject
 
{\bf Figure 1:} 
Color-composite images of the region of the HDF-N near the host of SN 1997ff.
The WFPC2 images were taken during the HDF-N campaign (Williams et al. 1996) and
the NICMOS images were taken during the GTO campaign (Thompson et al. 1999).
The arrow indicates the SN host galaxy.

{\bf Figure 2:}
{\it HST} exposures obtained for SN 1997ff as a function of time in different
bandpasses.  Time in days (the abscissa) is given relative to December 23,
1997, the start of the HDF-N SN search with WFPC2 (GO 6473).  The subsequent GTO
campaign with NICMOS (Thompson et al. 1999; GTO 7235) and its preceding tests
for guide-star suitability (7807) provided valuable coverage of the light curve
of the SN found during the search.  A subsequent NICMOS campaign (Dickinson et
al. 2001; GO 7817) provided templates to remove the contaminating light of the
host after the SN had faded.

{\bf Figure 3:} 
SN 1997ff in F814W ($I$), F110W ($J$), and F160W ($H$).  The images on the left
show the region of the HDF-N near the SN host without the SN (template images).
The images on the right show the difference in intensity between a SN image and
the template image.  Superimposed on this image are intensity contours.
Spectroscopic redshifts (Cohen et al. 2000) are listed as exact while
photometric redshifts (Budav\'ari et al. 2000) are listed as approximate.

{\bf Figure 4:}
Photometric redshift estimate for the host of SN 1997ff.  This estimate employs
6 photometric magnitudes from {\it HST} observations and 3 from ground-based
observations (Budav\'ari et al. 2000).  The galaxy magnitudes are given in
Table 1.  The top panel shows the best fits for the CWW (galaxy SEDs), KL2
(eigenspectra), and KL5 (eigenspectra) models.  The middle plot shows the
sensitivity of the fit to the redshift.  The bottom plot shows the biases which
result from using only the WFPC2 data.

{\bf Figure 5:} 
A comparison between the observed and expected near-IR colors of a SN Ia as a
function of assumed redshift.  For a given redshift, the color evolution is
plotted beginning 15 days before maximum (bluest point) and again in 5-day
intervals.  SNe Ia redden with time, reach their reddest color at 25--30 days
after maximum, and then become bluer by a few tenths of a magnitude.  The
relative size of the point scales with temporal proximity to $B$ maximum.  The
observed colors of SN 1997ff at discovery and $30/(1+z)$ days later puts
strong constraints on the redshift and age of the SN.

{\bf Figure 6:}
Optical and near-IR spectroscopy of the host, HDF 4-403.0, from the Keck
telescope.  The upper panel shows the optical spectra of the SN host compared
to the spectrum of an old, red elliptical galaxy, LBDS 53w091 ($z=1.55$;
Spinrad et al. 1997) transformed to $z=1.755$.  A simple $\chi^2$ minimization
provides a possible match at a redshift for the SN host of $z=1.67$ to 1.79,
but this match is not robust.  The bottom panel shows the near-IR
spectroscopy of the host.  An apparent weak emission line, if identified as 
[O~II] $\lambda$3727, would yield $z=1.755$, but this redshift determination is
tentative.  A gradient in the detected continuum is apparent, with an increase
to the red.

{\bf Figure 7:}
Comparison between the $B$-band light curve of a normal SN Ia and the observed
data transformed to rest-frame $B$ for different assumed redshifts and
discovery ages.  The observed SN colors, or, for the transformation to a fixed
bandpass shown here, the $K$-corrections are a strong function of redshift and
SN age.  The distance modulus may be constrained by offsetting the model light
curve in magnitudes.  A good fit between model and data only occurs in a narrow
range of redshifts and ages as seen in the middle panel.

{\bf Figure 8:} 
Marginalized probability density functions for SN Ia model parameters used to
fit SN 1997ff.  The top panel shows the PDF for the redshift constraints from
the galaxy photometry, from the SN, and from both sources.  The middle panel
shows the constraint on the age of the discovery relative to $B$ maximum.  The
bottom panel shows two forms of the observed luminosity function for SNe Ia
and the constraints on the luminosity of SN 1997ff using the priors and
the light-curve fits.

{\bf Figure 9:}
Confidence intervals for the discovery age (days relative to $B$-band maximum)
and redshift of SN 1997ff.  Because the observed SN colors are a strong
function of both of these parameters, a high degree of correlation exists in
their simultaneous determination.

{\bf Figure 10:}
Confidence intervals for the distance modulus $(m-M)$ and redshift of SN
1997ff.  The intervals were calculated using the galaxy photometric redshift
and the observed luminosity function of SNe Ia (optimal), neglecting the galaxy
photometric redshift, and using a weak prior on SN Ia luminosities described in
the text.

{\bf Figure 11:} 
Hubble diagram of SNe Ia minus an empty (i.e., ``empty'' $\Omega=0$) Universe
compared to cosmological and astrophysical models.  The points are the
redshift-binned data from the HZT (Riess et al. 1998) and the SCP (Perlmutter
et al. 1999).  Confidence intervals (68\%, 95\%, and 99\%) for SN 1997ff are
indicated.  The modeling of the astrophysical contaminants to cosmological
inference, intergalactic grey dust or simple evolution, is discussed in \S 4.2.
The measurements of SN 1997ff are inconsistent with astrophysical effects which
could mimic previous evidence for an accelerating Universe from SNe Ia at
$z \approx 0.5$

{\bf Figure 12:}
Same as Figure 11 with the inclusion of a family of plausible, flat
$\Omega_\Lambda$ cosmologies.  The transition redshift (i.e., the coasting
point) between the accelerating and decelerating phases is indicated and is
given as $ \left[ 2\Omega_\Lambda /\Omega_M \right]^{1/3} -1$.  SN 1997ff is
seen to lie within the epoch of deceleration.  This conclusion is drawn from the
result that the apparent brightness of SN 1997ff is inconsistent with values of
$\Omega_\Lambda \geq 0.9$ and hence a transition redshift greater than that of
SN 1997ff.

{\bf Figure 13:} 
SN type versus host morphology as compiled by the Asiago catalog (Cappellaro et
al. 1997).  This set includes all SNe through SN 2001X for which a modern SN
classification and galaxy classification are available.

{\bf Figure 14:}
Possible star-formation histories of the host of SN 1997ff compared to its
observed SED.  The upper panel shows the expected SED for a single burst of
star formation occurring at 0.5 Gyr, 1 Gyr, and 2 Gyr before the explosion SN
1997ff.  The favored age of 1 Gyr is long after the loss of the progenitors of
core-collapse SNe.  For a star-formation history consisting of a burst (lower
panel) followed by continuous and exponential decay of extended star formation
($\tau = 0.3$ Gyr), the expected ratio of SNe Ia to core-collapse SNe is
between 20 and 70.

{\bf Figure 15:}
The tangential ellipticity test for the SN host galaxy.  The points show a
subset ($10^4$ of the $10^6$ samples) of randomly selected combinations of
the intrinsic axis ratio and orientation angle of simulated host galaxies.  For
several different hypothetical values of the SN amplification, a curve is shown
indicating the combinations of intrinsic axis ratio and orientation angle
resulting in the observed tangential eccentricity of the SN host.  Each curve
is labelled with a hypothetical amplification (in magnitudes) of the SN and the
fraction of randomly selected galaxies whose observed tangential ellipticity
would be as small as that observed or smaller (see Appendix C).

\vfill \eject

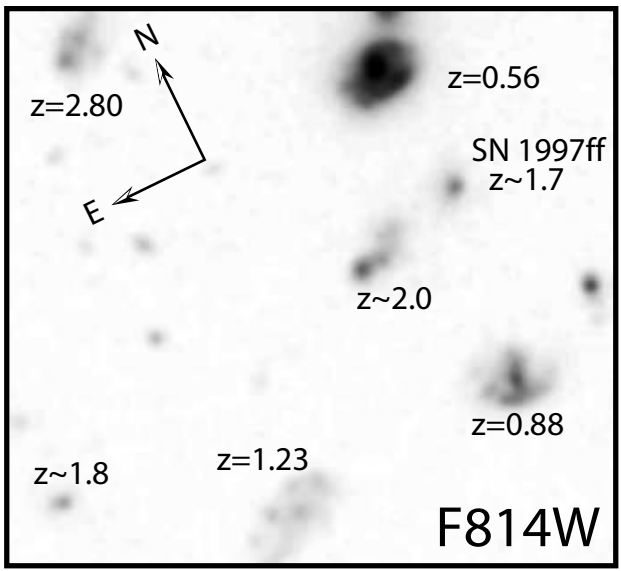
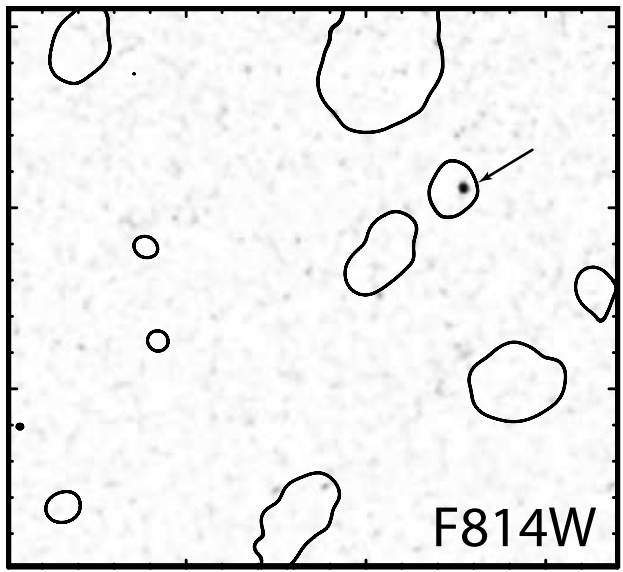
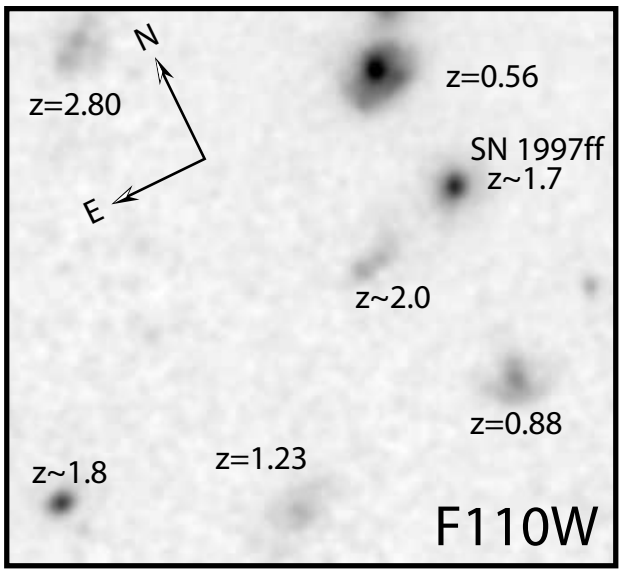
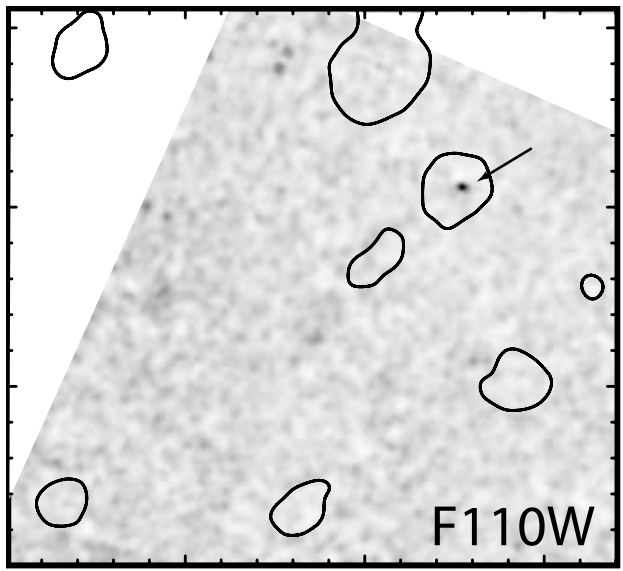
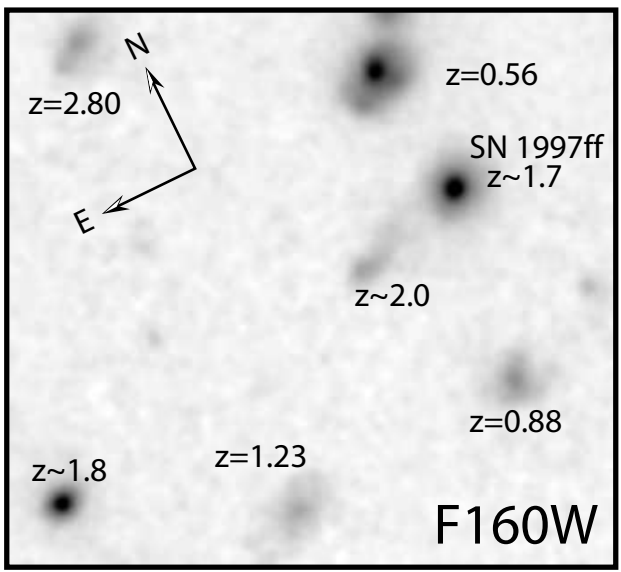
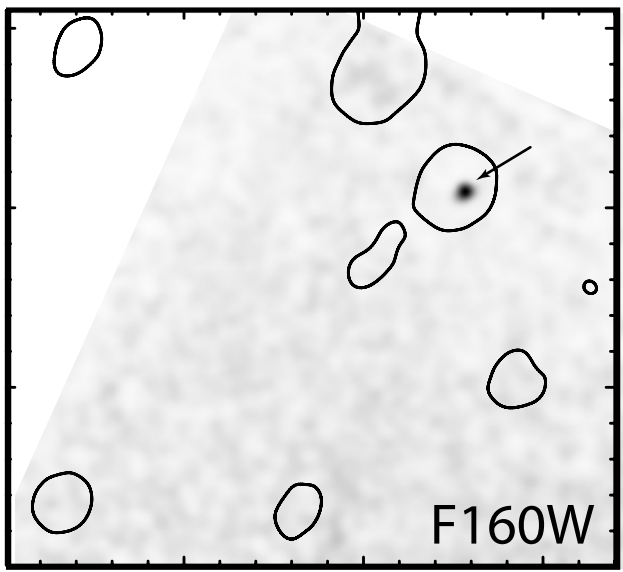
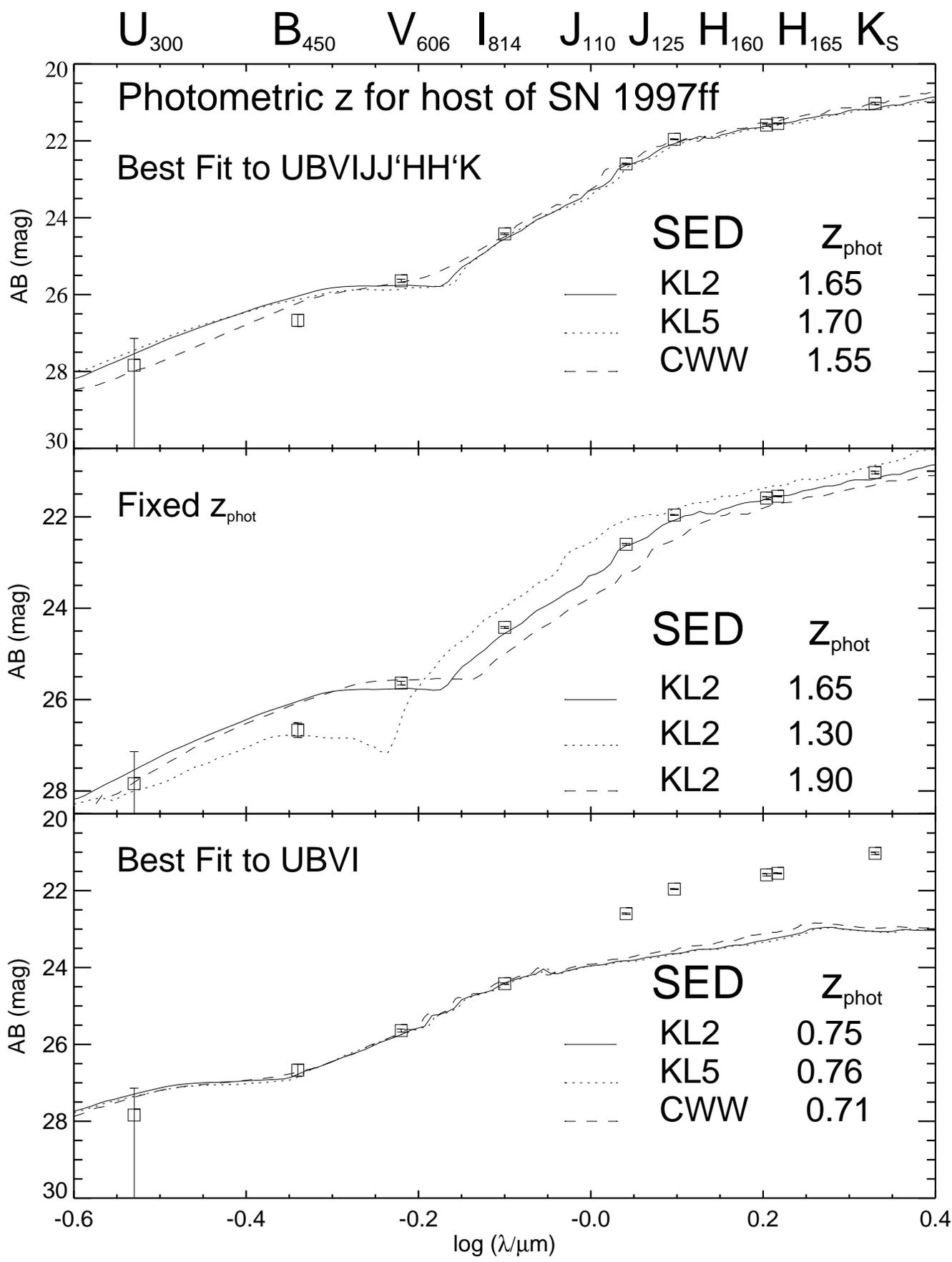
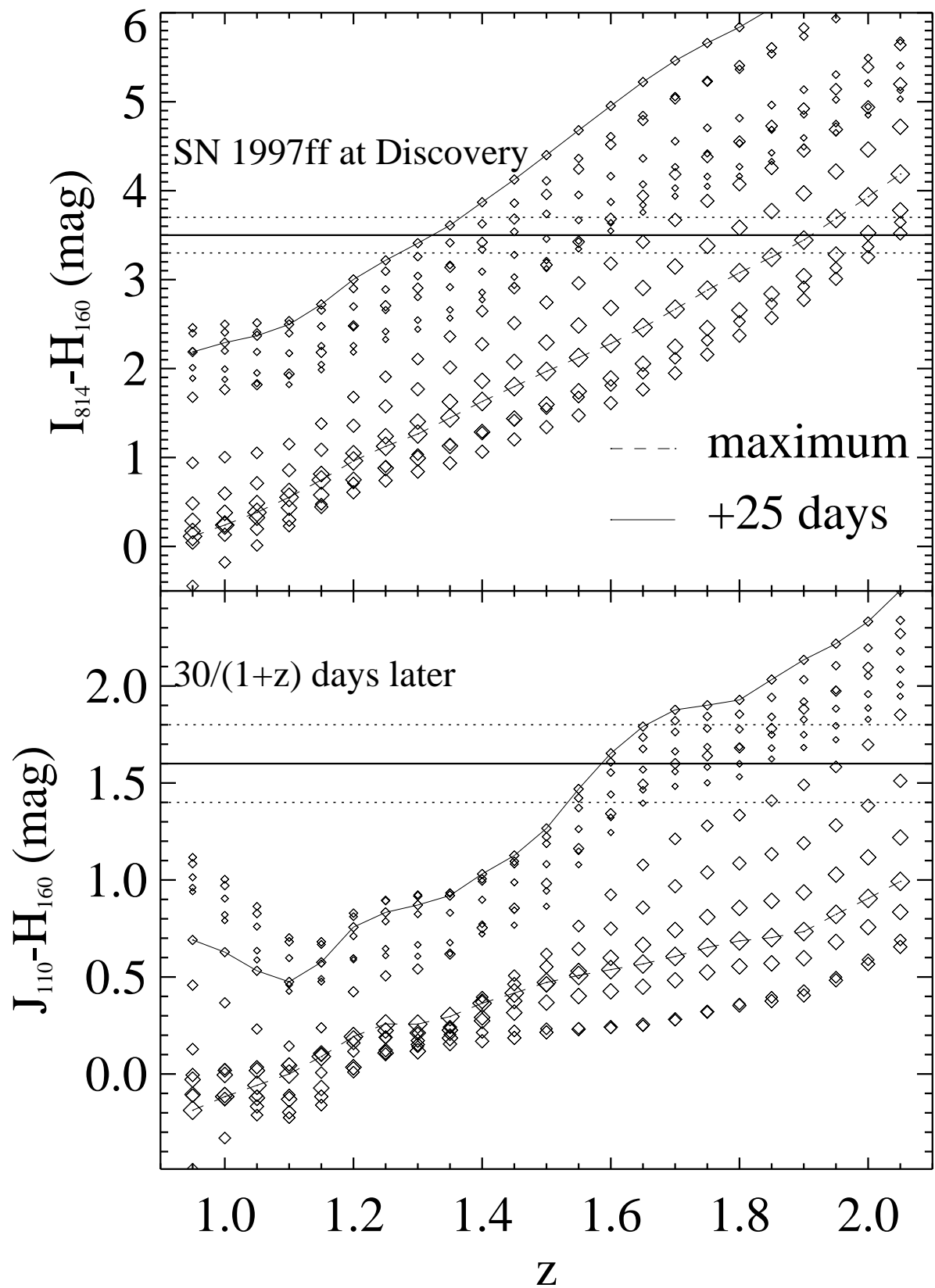
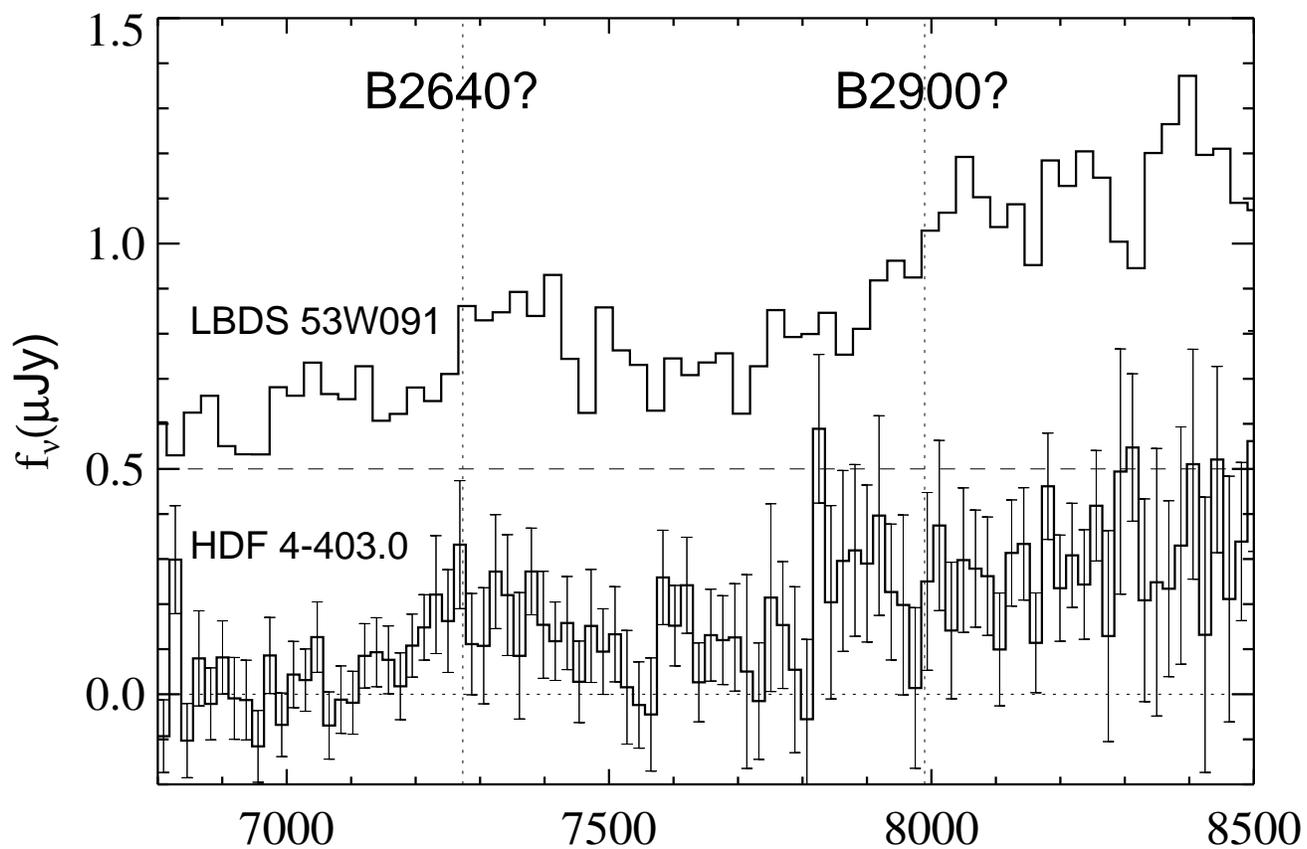
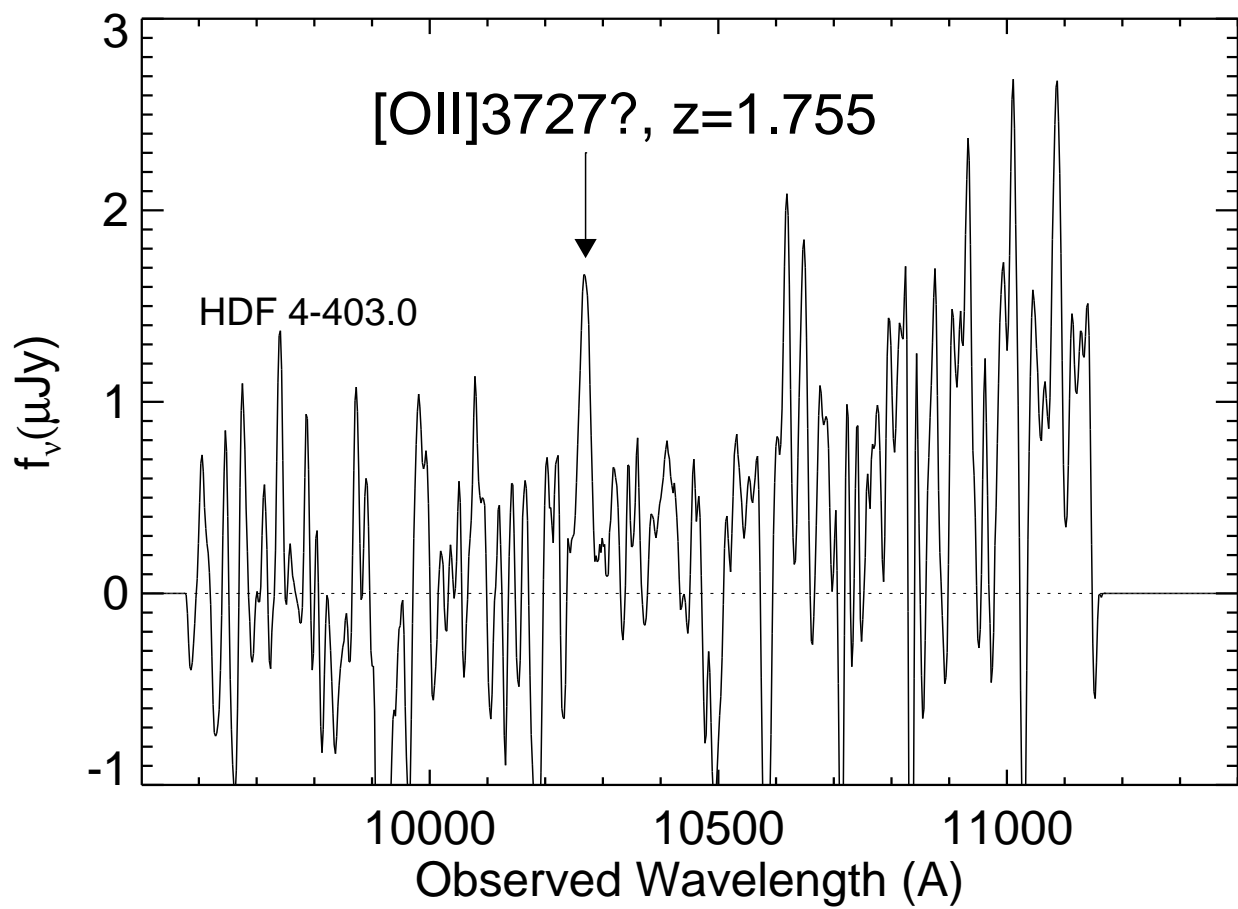
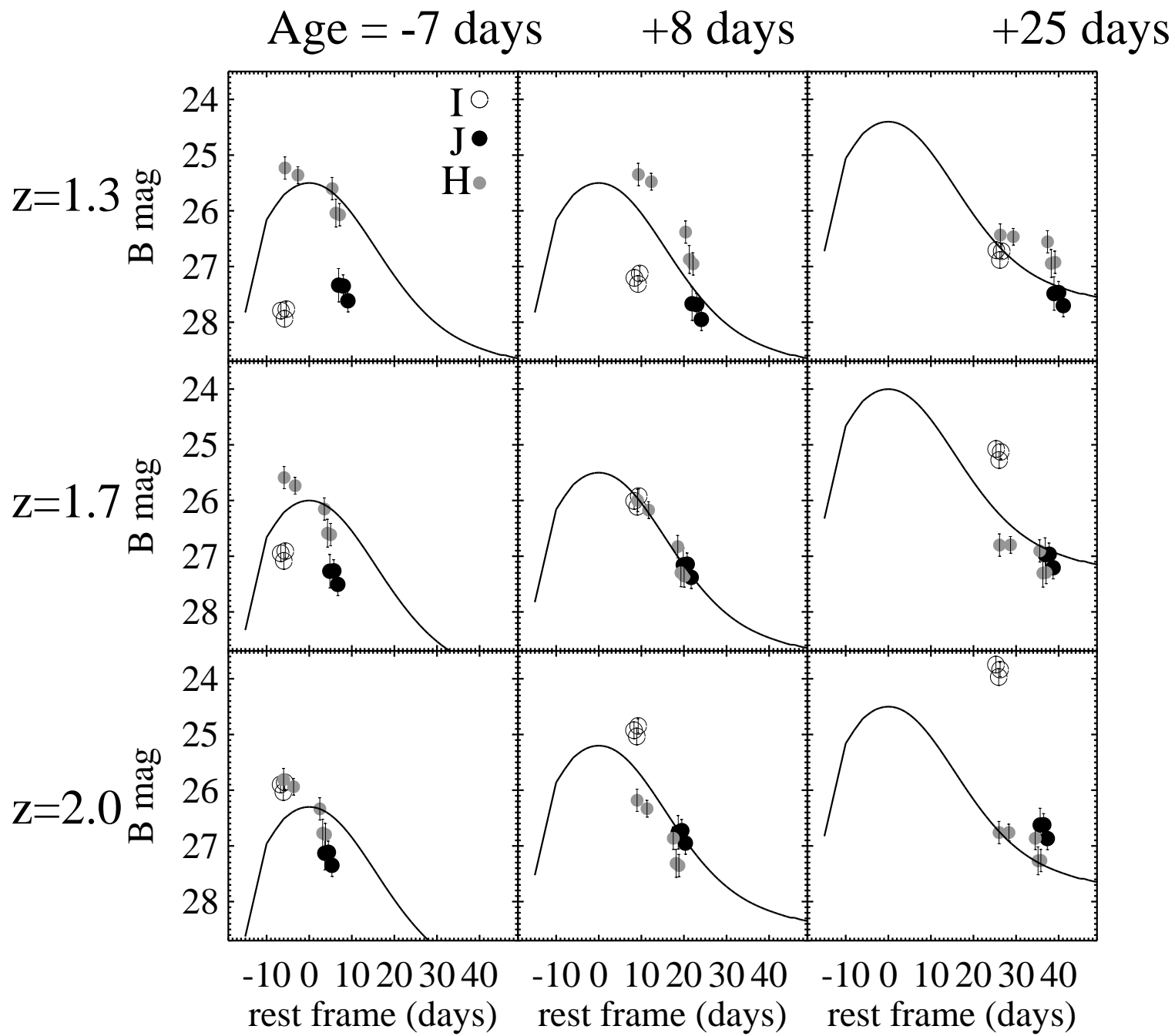
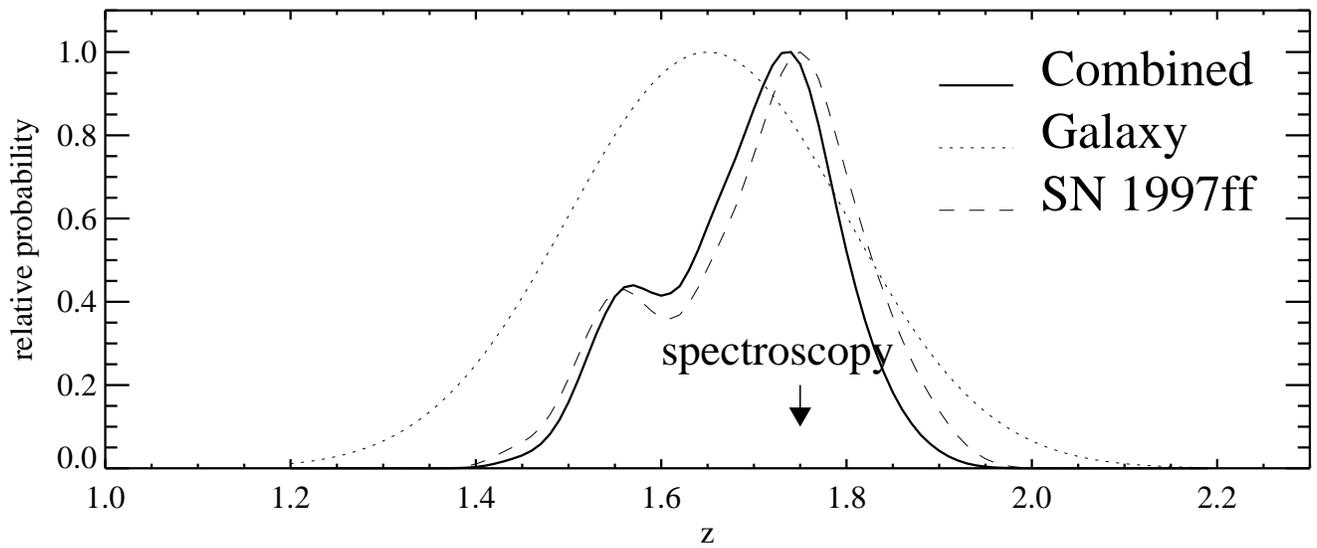
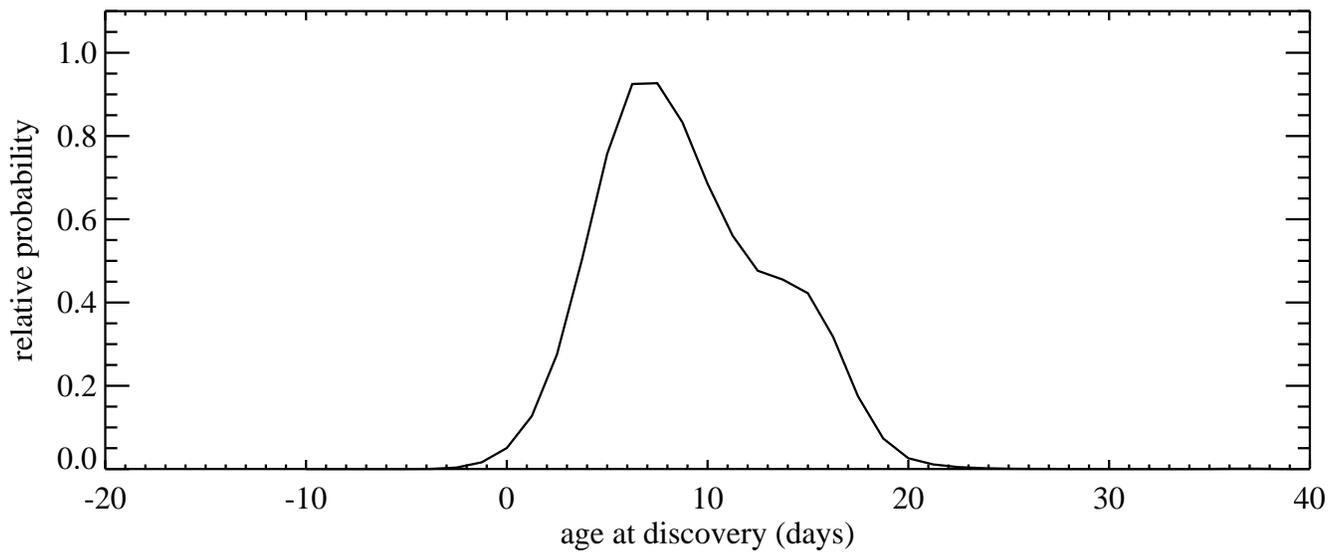
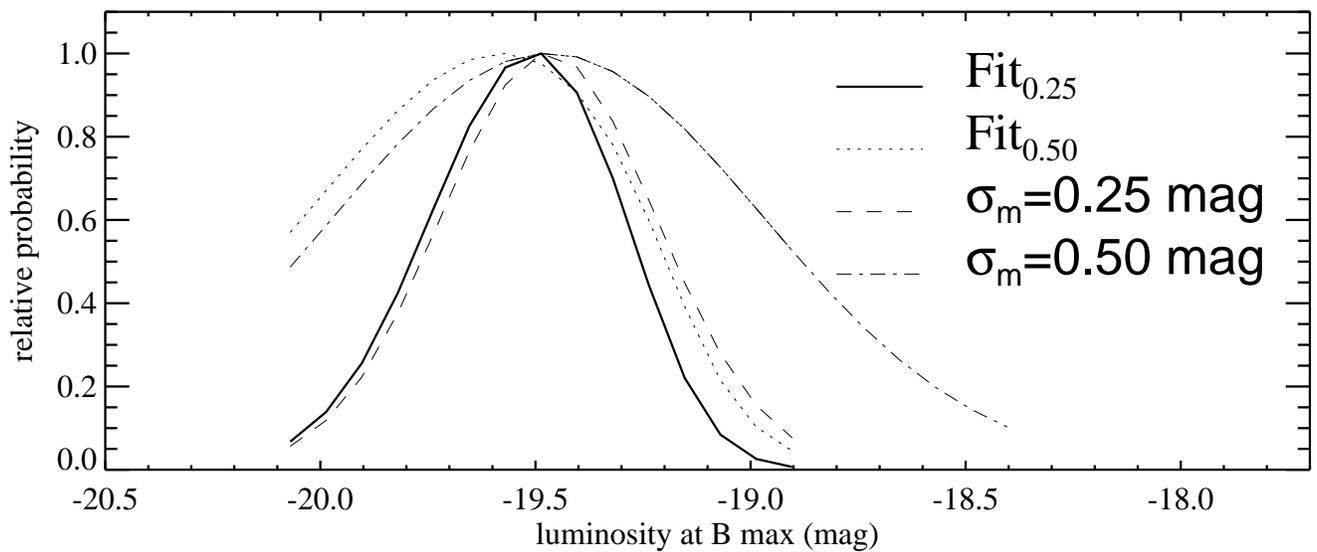
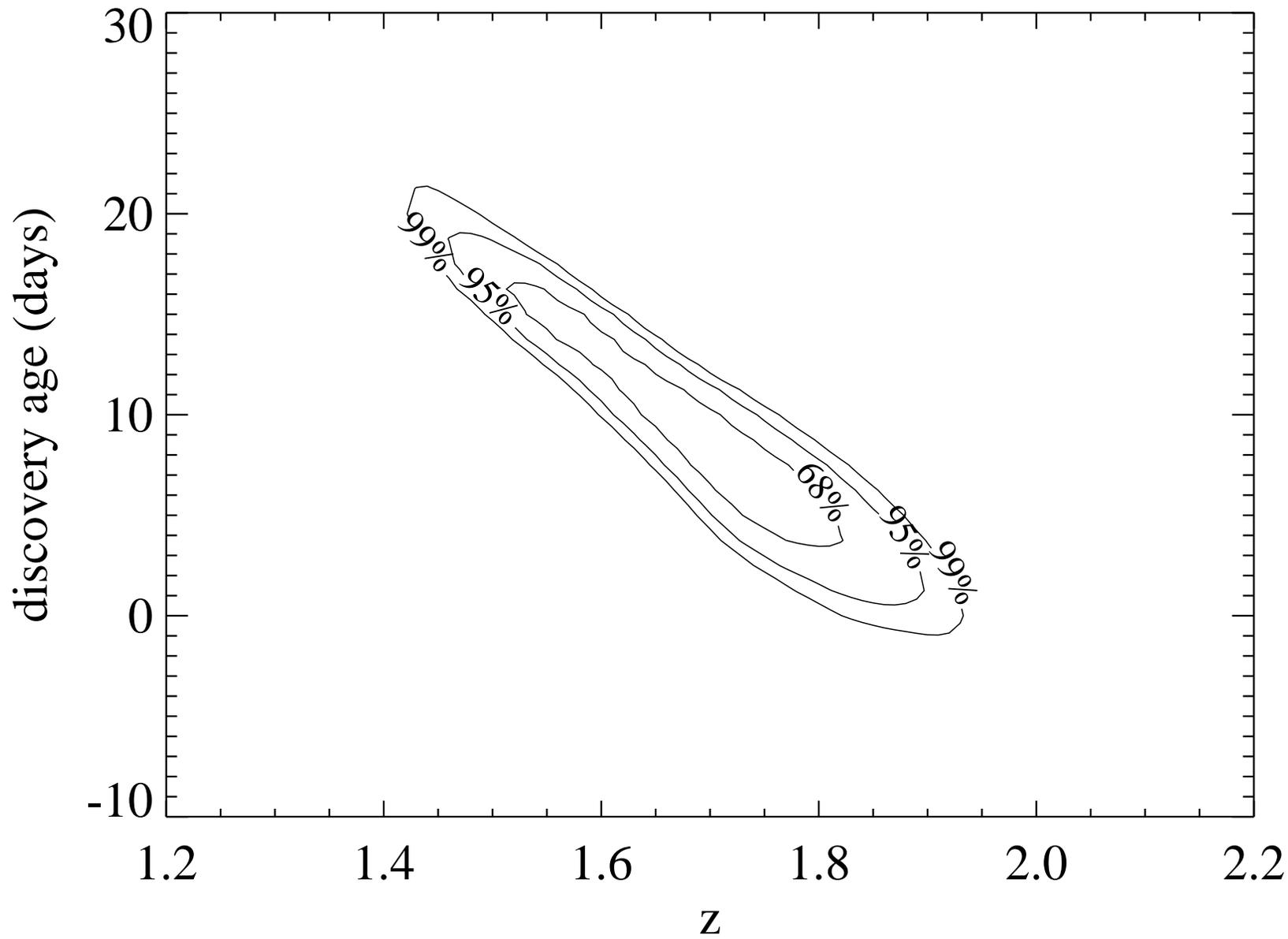
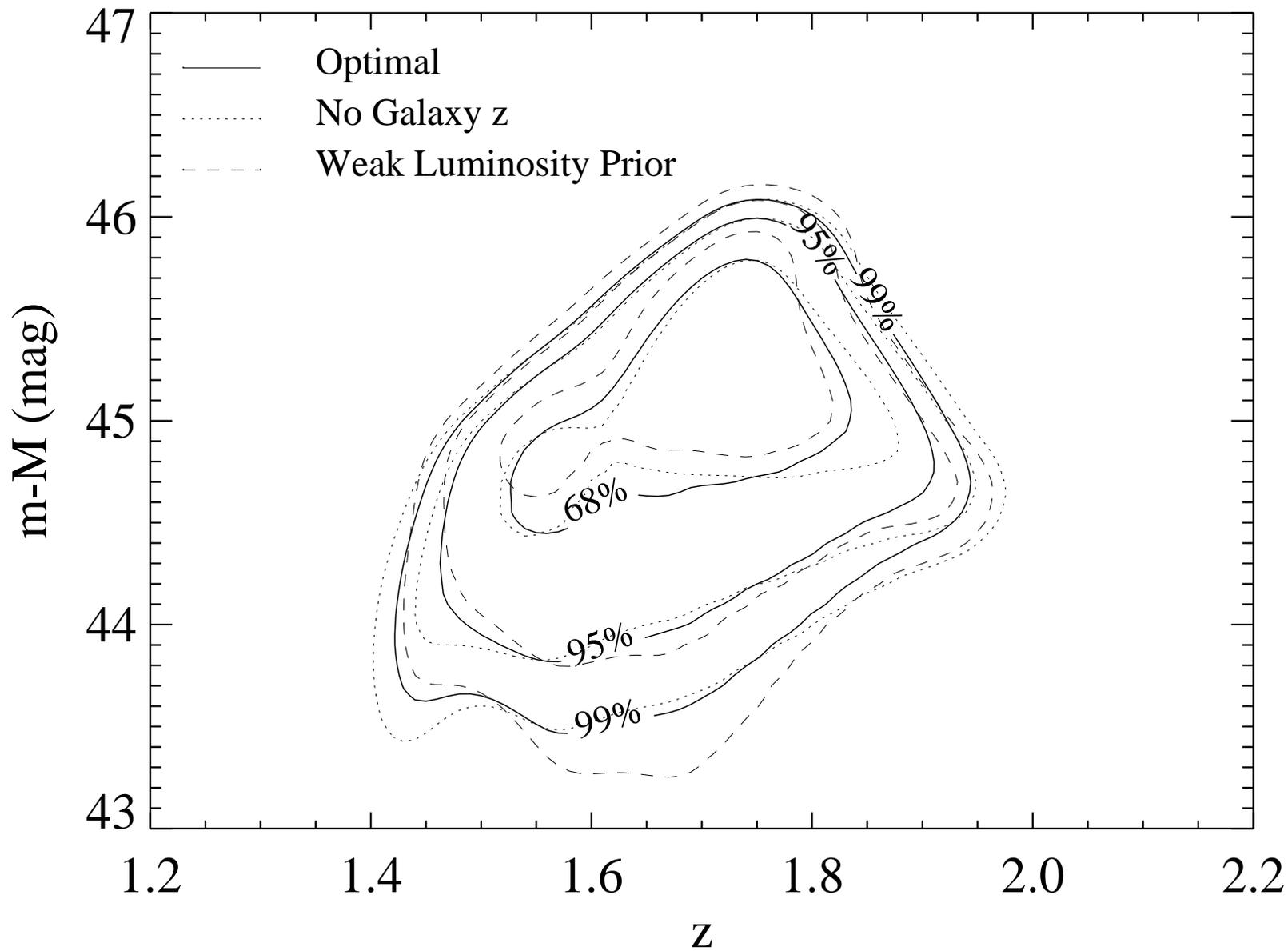
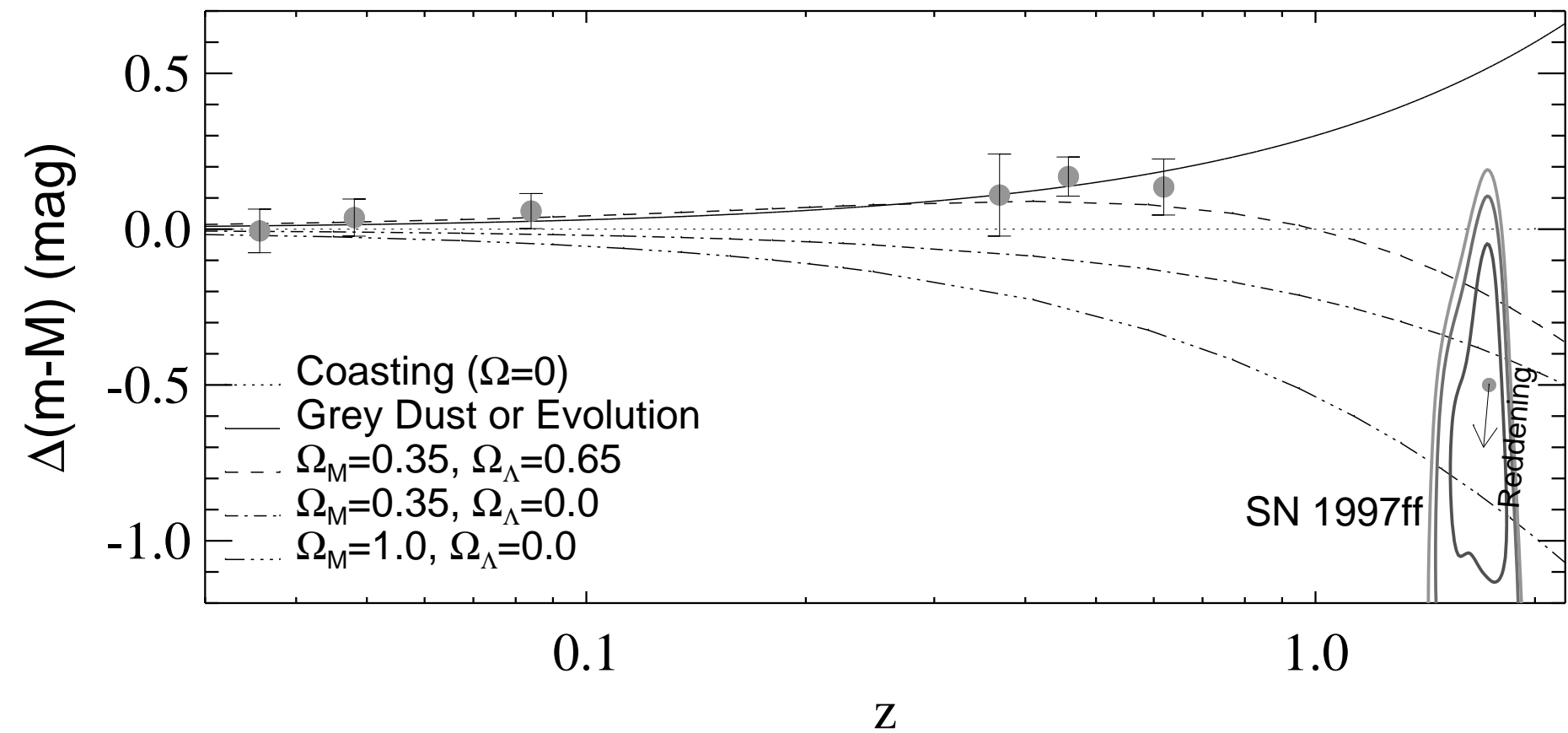
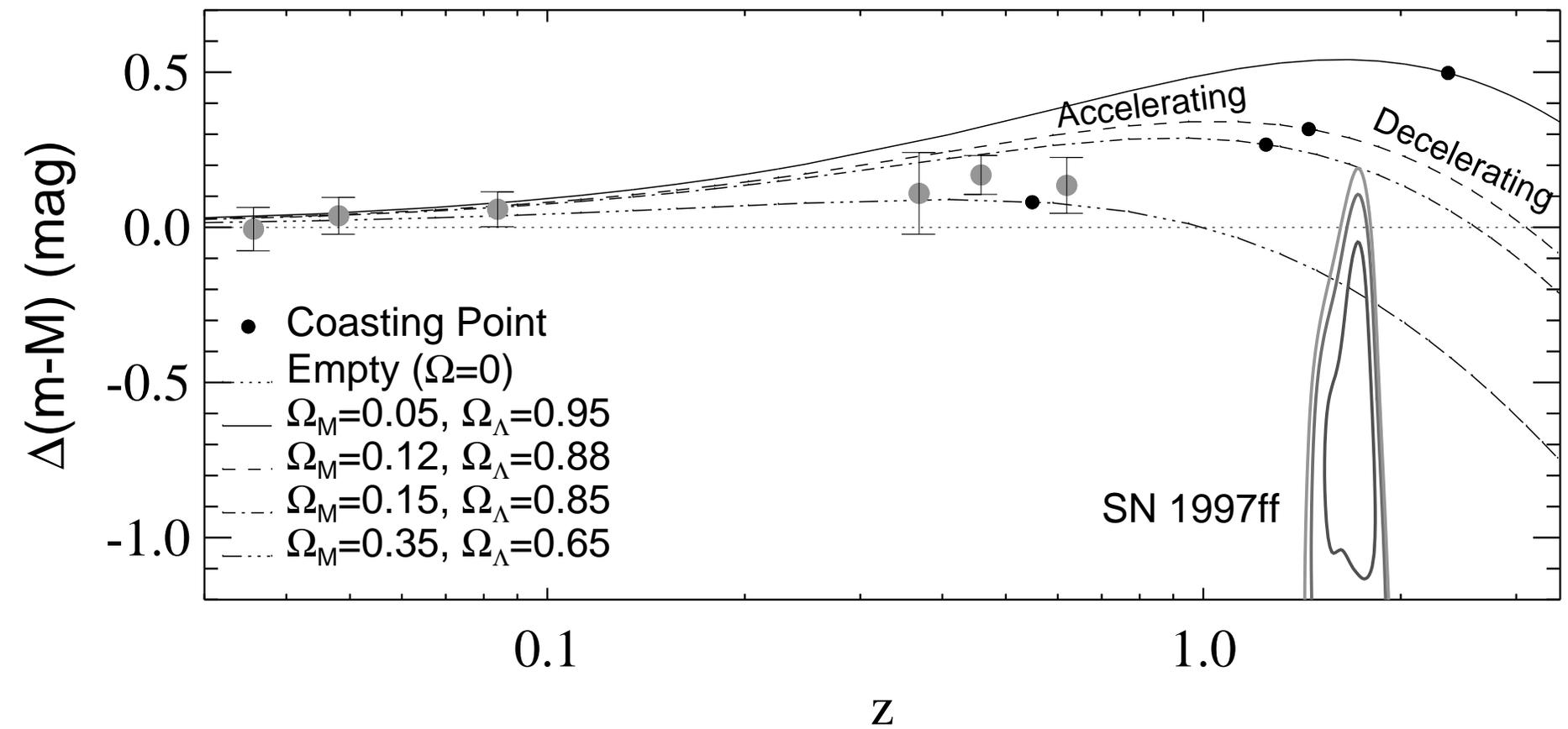
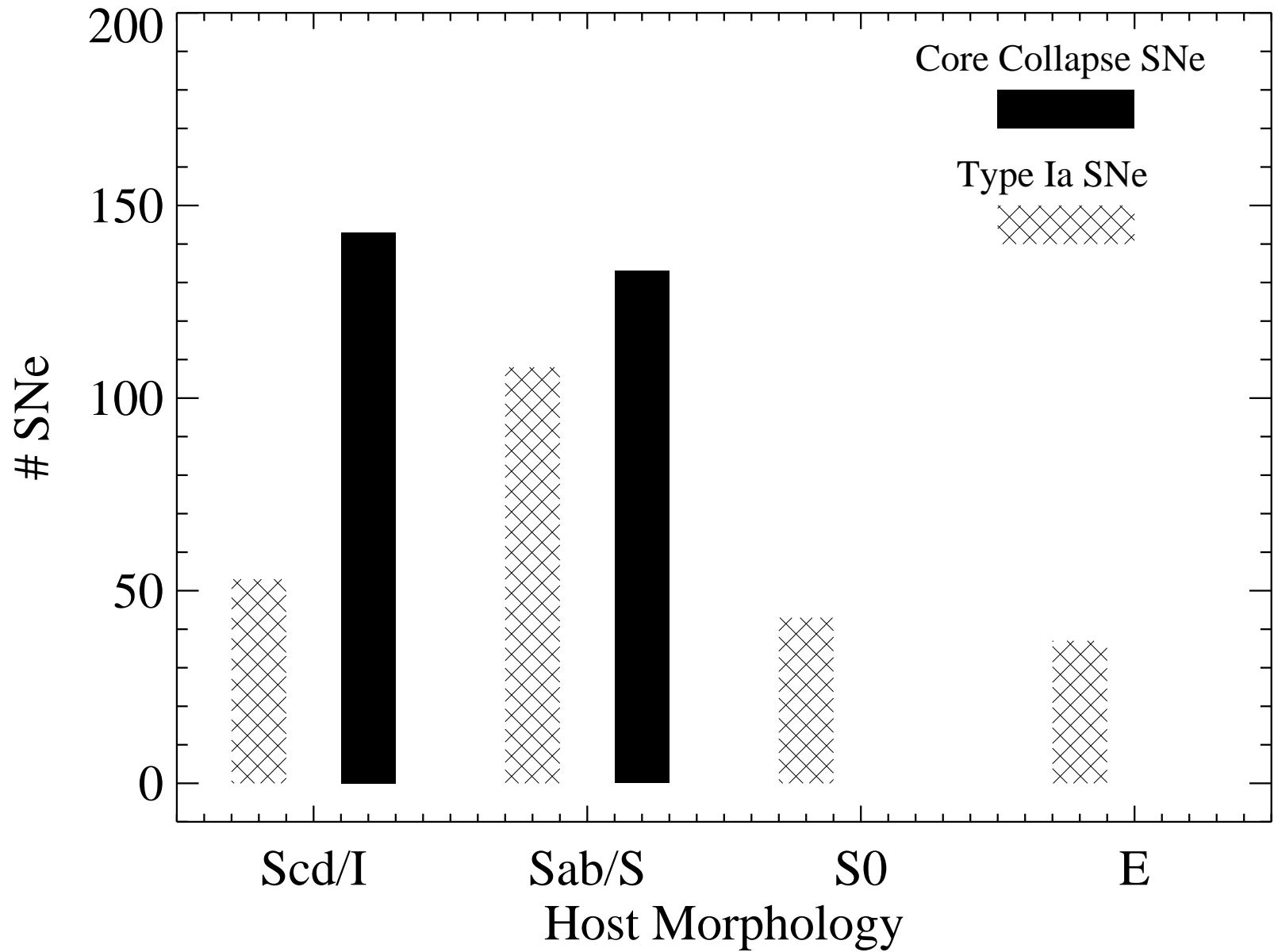
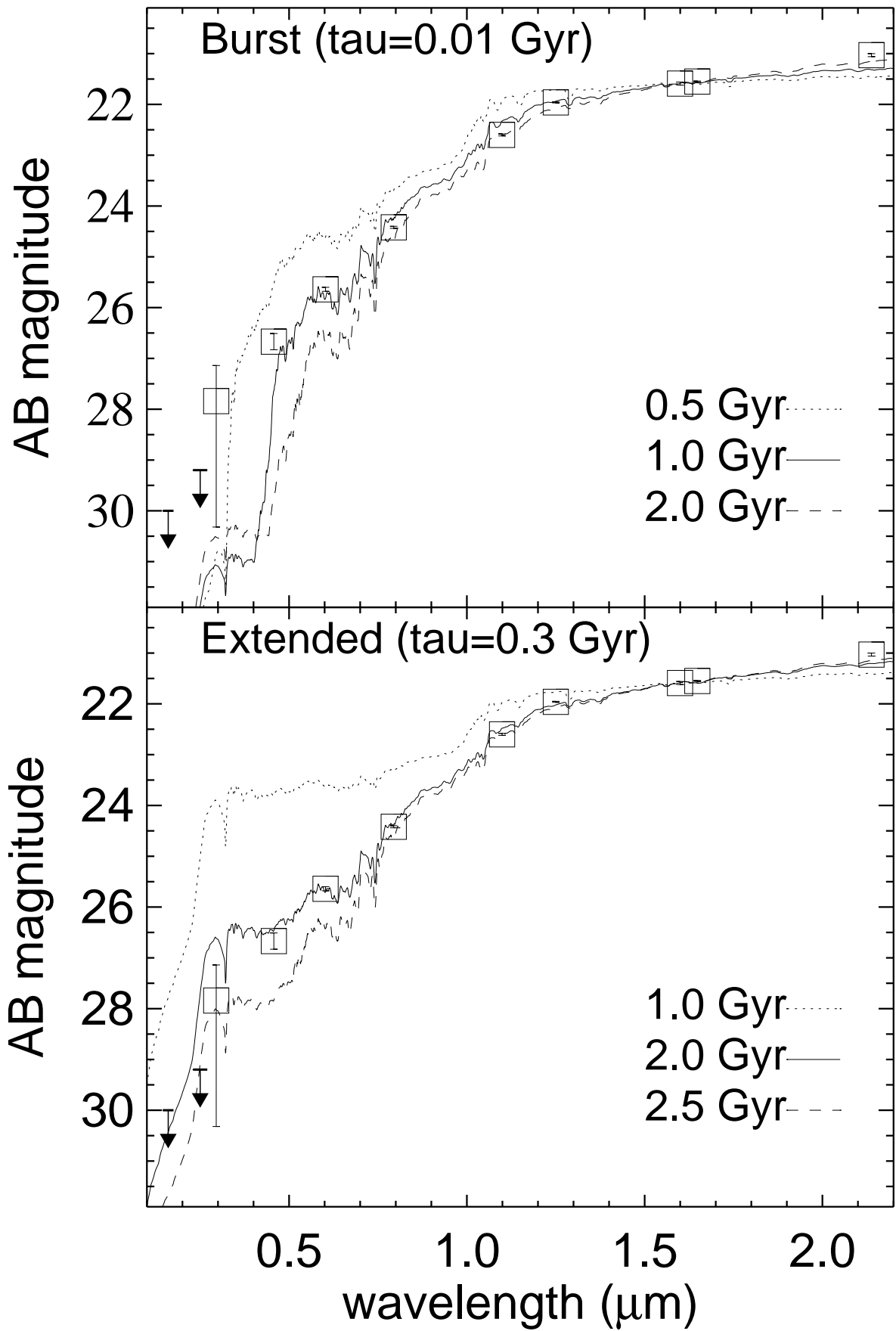
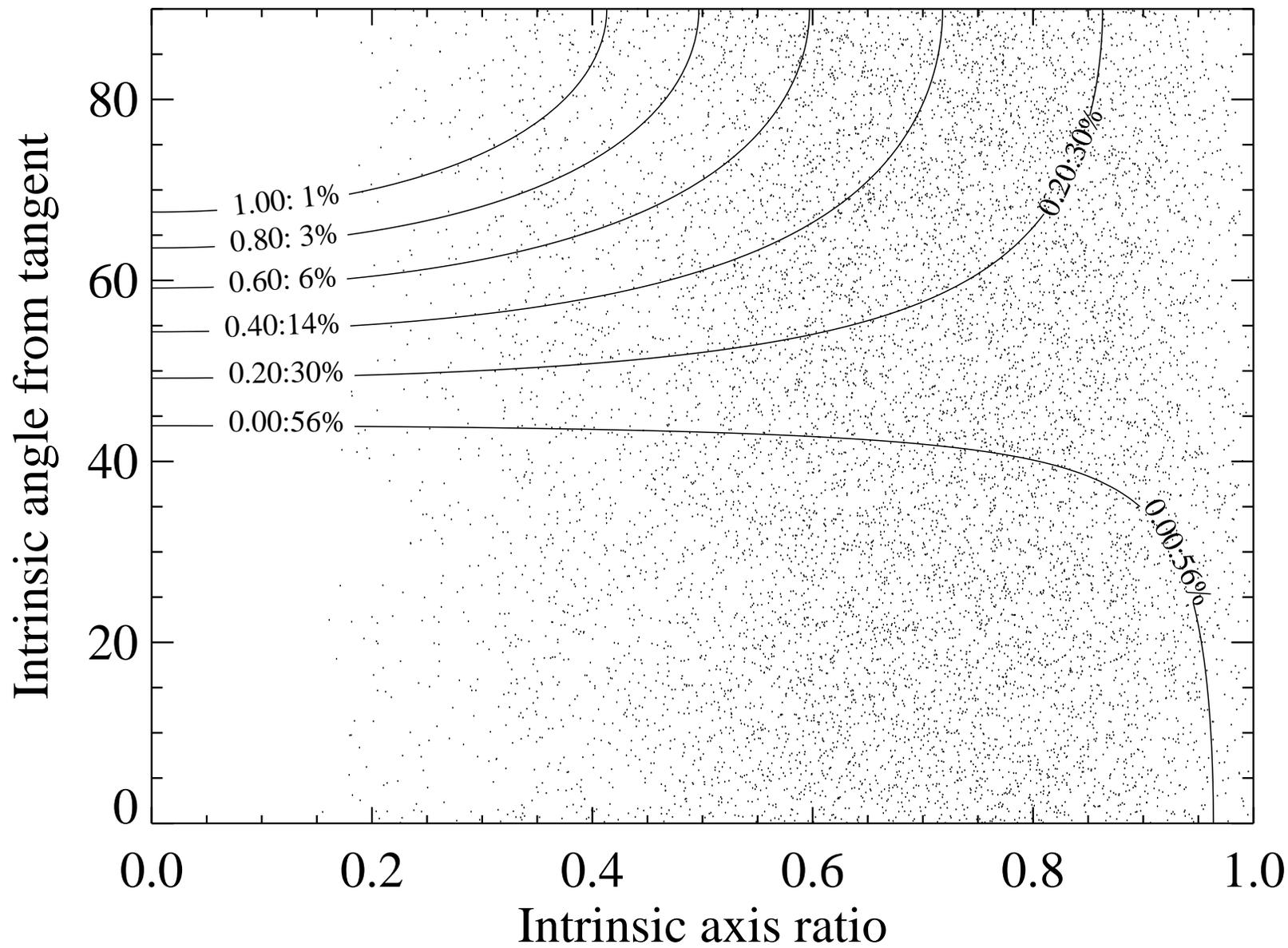
\end{document}